\documentclass[pre,aps,superscriptaddress,twocolumn]{revtex4}

\usepackage{graphicx}
\usepackage{amsmath,amsfonts}

\newcommand{\beq}{\begin{equation}}
\newcommand{\eeq}{\end{equation}}
\newcommand{\bea}{\begin{eqnarray}}
\newcommand{\eea}{\end{eqnarray}}

\begin{document}
\title{Topological properties of the mean field $\phi^4$ model}
\author{A.~Andronico}
\affiliation{Dipartimento di Fisica and INFM, Universit\`a di Roma {\em La Sapienza},
  P. A. Moro 2, 00185 Roma, Italy}
\author{L.~Angelani}
\affiliation{Dipartimento di Fisica and INFM, Universit\`a di Roma {\em La Sapienza},
  P. A. Moro 2, 00185 Roma, Italy}
\affiliation{INFM - CRS SMC
  Universit\`a di Roma {\em La Sapienza}, P. A. Moro 2, 00185 Roma, Italy}
\author{G.~Ruocco}
\affiliation{Dipartimento di Fisica and INFM, Universit\`a di Roma {\em La Sapienza},
  P. A. Moro 2, 00185 Roma, Italy}
\affiliation{INFM - CRS Soft
  Universit\`a di Roma {\em La Sapienza}, P. A. Moro 2, 00185 Roma, Italy}
\author{F.~Zamponi}
\affiliation{Dipartimento di Fisica and INFM, Universit\`a di Roma {\em La Sapienza},
  P. A. Moro 2, 00185 Roma, Italy}

\begin{abstract}
We study the thermodynamics and the properties of the stationary points (saddles and minima)
of the potential energy for a $\phi^4$ mean field model.
We compare the critical energy $v_c$
(i.e. the potential energy $v(T)$ evaluated at the phase transition temperature $T_c$)
with the energy $v_\theta$
at which the saddle energy distribution show a discontinuity in its derivative.
We find that, in this model, $v_c \gg v_\theta$,
at variance to what has been found in the literature for different mean field and short
ranged systems.
By direct calculation of the energy $v_s(T)$ of the ``inherent saddles'', i.e. the saddles visited by the
equilibrated system at temperature $T$, we find that $v_s(T_c) \sim v_\theta$.
Thus, we argue that the thermodynamic phase transition is related to a change in the properties of
the inherent saddles rather then to a change of the topology of the potential energy surface at
$T=T_c$.
Finally, we discuss the approximation involved in our analysis and the generality of our method.
\end{abstract}

\maketitle

%%%%%%%%%%%%%%%  TEXT  %%%%%%%%%%%%%%%%

\section{Introduction}

The investigation of the topological properties of the potential
energy surface (PES) of liquids and disordered system
\cite{stillinger}, has been strongly revitalized in the last years
\cite{deb_nature,land_angell,land_sastry,land_buc,land_sastry2,land_keyes,noiptrig}.
These studies have been
particularly focused on the connection between the slow dynamics
of supercooled liquids and the properties of the stationary points
of the potential energy function $V(q)$, being
$q_i$ ($i=1,\cdots,N$) the set of $N$ generic configurational variables.

In the first approaches, on studying the slow dynamics of
supercooled liquids and glasses, the objects of the investigations
were the properties --energy location ($v_m$), curvature
($\omega_m$), etc.-- of the minima of the PES that are ``visited'' by
the system during its evolution at a given thermodynamic state.
Assigning to each minimum its basin of attraction, one obtains a
partition of the configurational phase space: to each
instantaneous configuration $q$, whose instantaneous potential
energy is $v=V(q)/N$, one associates an ``inherent'' configuration
$q_m$, whose potential energy is $v_m=V(q_m)/N$. This
allows one to define a configurational entropy of the minima and a
free energy for the supercooled and for the out-of-equilibrium
glassy regime~\cite{fs_entropy}. These properties of the minima of
the PES have then been connected to several features of
supercooled liquids and glasses. Among them, we mention
the {\it fragility} of the glass former~\cite{cav-fragility,land_sastry2,noifragility},
the diffusion processes in supercooled
liquids~\cite{land_keyes,fabr,la_98,donati}, and the effective
fluctuation-dissipation
temperature~\cite{FDTgen} in the
out-of-equilibrium glassy phase~\cite{fs_aging}.

More recently this minima-based approach has been extended to
consider also the other stationary points of the PES, namely the
saddle points. Using the saddle-based approach, it has been
shown in Lennard-Jones like liquids \cite{noi_sad,cav_sad} and in
p-spin mean field systems \cite{cav-pspin} that
the ``order of the inherent saddles'' (i.e. the number of negative
eigenvalues of the Hessian matrix evaluated at the saddle points
visited during the equilibrium dynamics at temperature $T$)
extrapolates to zero when $T$ reaches the dynamic transition
temperature $T_{MCT}$ (or mode-coupling
temperature~\cite{mct}).
While the definition of ``basin of attraction of a saddle'' and the
operative way to associate a saddle point to the instantaneous
configuration of the system --i.e. the way to associate a saddle
$q_s$ (with energy $v_s=V(q_s)/N$) to each instantaneous configuration $q$
(with energy $v$)-- have been the subject of debate
\cite{sad_1,doye,sad_3}, the previous result has been show to be
robust and the method has been applied to other model systems
\cite{sad_3,sadBLJ,parisi_boson,termoselle}.
In the following we will call a ``map'' the function that associate the
thermal average of $v_s=V(q_s)/N$ to its parent $v=V(q)/N$, i.e.
for each $T$, if $v(T)$ is the average potential energy and
$v_s(T)$ the average potential energy of the saddle visited at
temperature $T$, then the map is the function $\cal{M}$ such that
$v_s(T)={\cal M}(v(T))$. Until now, two different operative
definitions of saddle to be associated to an instantaneous
configuration (two different maps) have been used: {\it 1)}~In the
numerical simulations of simple models, such as Lennard-Jones
systems, a partitioning of the configuration space in basins of
attraction of saddles is obtained via an appropriate function
$W(q)$ (usually $W(q)=|\nabla_q V(q)|^2$) that has a
local minimum on each stationary point of $V(q)$, and the
saddles are then obtained via a minimization of $W(q)$
starting from an equilibrium configuration obtained from a
molecular dynamics simulation at temperature $T$ \cite{noi_sad,cav_sad}.
{\it 2)}~In the analytic computations applied to disordered mean-field spin models
one looks to the saddles that are closest, with respect to the
distance in the configuration space, to a reference configuration
extracted from the Gibbs distribution at temperature
$T$~\cite{cav-pspin}. In one specific case, the only one where this test
has been performed, the two definitions have been proven to give
identical results \cite{noiptrig}.

The role of the stationary points of the PES (saddles and minima)
has been also pointed out in a different context. Recent studies
aiming to clarifying the microscopic origin of phase transitions
suggest that the presence and order of such transitions are related
to changes in the topology of the manifold of the PES sampled by
the system when crossing the (thermodynamic) critical point. More
specifically the topology change is signaled by a discontinuity in
(one of) the derivatives of the Euler characteristic. This
function, determined by counting the number and the order of the
stationary points of the PES, is a genuine topological property of
the constant potential energy sub-manifold, and, in particular, it
does not depend on the statistical measure defined on it (i.e., on
temperature).

Before to proceed, it is worth to observe that the Euler
characteristic $\chi(v)$ (that is used in the topological studies
of the phase transitions) and the complexity of minima and saddles
(which is the basic quantity in the investigation of the role
played by the stationary points of the PES in determining the slow
dynamics in disordered systems) have {\it similar} definitions.
Being ${\cal N}_\nu(v)$ the number of stationary points $q_s$ of order $\nu$
(minima for $\nu$=0 and saddles for $\nu\geq$1) that have potential
energy $V(q_s) \leq N v$, we can define the energy distribution of the
stationary points
\begin{equation}
\label{complexity}
\Omega(v) = \sum_\nu {\cal N}_\nu(v) \ ,
\end{equation}
and the Euler characteristic
\begin{equation}
\label{eulerdef}
\chi(v)= \sum_\nu (-1)^\nu {\cal N}_\nu(v) \ .
\end{equation}
It is clear that these two definitions are similar, but not
coincident. Specifically, as ${\cal N}_\nu(v)$ is usually
exponentially large in the size of the system $N$, $\Omega(v)$ can
be evaluated at the saddle point in $\nu$, thus defining an order
$\bar\nu(v)$ that dominates in Eq.~(\ref{complexity}), while this
procedure may not apply to $\chi(v)$ where large cancellations can
arise from the term $(-1)^\nu$. The complexity (or configurational
entropy) $\sigma(v)$ is defined as the logarithm of the number of
stationary points whose energy lies in $[v,v+\delta v]$:

\beq \sigma(v) = \frac{1}{N} \log \left[ \frac{d\Omega(v)}{dv}
\delta v \right] \sim \frac{1}{N} \log \Omega(v) \eeq

where the last approximation is promptly obtained recalling that
$\Omega(v) \sim \exp N \sigma(v)$ is exponentially large in $N$.
This scaling is not always found for the Euler characteristic
that, at variance with $\sigma(v)$, can scale with $N$ in many
different ways \cite{phi42d}. We will further discuss this point
in the following.

Following the numerical results obtained in \cite{phi42d} on the
$\phi^4$ model with nearest neighbores interactions in two and
three dimensions, a theorem that relates the topological
properties of the PES to the thermodynamic phase transitions has
been recently demonstrated by Franzosi and Pettini for systems
with generic short range interactions \cite{TeoremaPettini}.
Thought the theorem strictly applies to non-mean field systems,
the mean field models examined so far seem to indicate the
existence of a topology-thermodynamics relation for mean field
systems as well. In the (mean field) XY model the (second order)
phase transition, that takes place at a temperature $T_c$ when the
system is visiting the PES level given by $v_c=v(T_c)$, is
signaled by a discontinuity in the first derivative of the Euler
characteristic $\chi(v)$ at $v=v_c$ \cite{PettiniXY}. In the
$k$-trigonometric model there is a phase transition, which is
second order for $k$=2 and first order for $k$$>$2. For all $k$
values, the phase transition is seen in the topology via a
discontinuity in the first derivative of $\chi(v)$ at $v_c$, and
the curvature of $\chi(v)$ around $v_c$ gives also information on
the order of the transition \cite{noieulero}. A detailed review of
the previous results can be found in~\cite{pettinisumma}.

To summarize the previous paragraph, it seems that the relation
between topology and thermodynamics is a general properties, being
demonstrated for short range systems and tested via explicit
computation of $\chi(v)$ for mean field system. There is, however,
a simple counterexample: the mean field $\phi^4$ model
\cite{russi,schillingDYN,tesiBaroni}. In Ref.~\cite{tesiBaroni} it
was observed that, for large value of the coupling parameters ($J$
in the following nomenclature), the phase transition (second
order, ferromagnetic-like) takes place at a temperature $T_c$
where the equilibrium potential energy value, $v_c$, is {\it
larger than the energy of the higher energy stationary point},
i.e. where $\chi(v)$=1 and, therefore, no discontinuity of
$\chi(v)$ can be present. At this stage of the discussion it is
worth to point out a feature that is common to all the mean field
cases (XY and k-trigonometric for any $k$) where the
topology-thermodynamics relation holds. Indeed, in these cases the
energy of the ``inherent'' saddle visited by the system at $T_c$
(i.e. $v_s(T_c)$) coincide with the instantaneous potential energy
$v(T_c)$. In other words, for these systems, $v_c$ is a fixed
point for the map ${\cal M}$: ${\cal M}(v_c)=v_c$. Thus, the
observed discontinuity of the derivative of $\chi(v)$ at $v_c$
cannot discriminate between the two possibilities: {\it i)}~is the
discontinuity in the topological properties {\it at the
instantaneous} potential energy that marks the phase transition,
or {\it ii)}~is the discontinuity in the topological properties
{\it at the inherent saddles} potential energy that marks the
phase transition. The $\phi^4$ model does not share this
peculiarity with the other investigated mean field models, and can
be therefore used to solve the ambiguity. While the
Franzosi-Pettini theorem for non-mean field system seems to favor
the possibility {\it i)}, the $\phi^4$ model indicates that {\it
i)} is not applicable in mean field systems. It is the aim of this
work to test whether the possibility {\it ii)} holds.

In this paper we first study the thermodynamics of the (symmetric)
$\phi^4$ model for different value of the coupling parameter $J$
(the only independent parameter of the model), in order to
individuate the temperature ($T_c$) and potential energy ($v_c$)
location of the second order ferromagnetic phase transition. We
then calculate the complexity of the stationary points of $V(q)$,
namely $\sigma(v)$, and we show that --at all $J$ values-- the
discontinuity of the derivative of $\sigma(v)$ is found at a value
($v_\theta$) which is always below $v_c$. Finally, we calculate
the energy $v_s$ of the inherent saddles (and thus the map
$v_s={\cal M}(v)$) in two different ways (minimization of $W(q)$
and lowest Euclidean distance), and we find that --within the
small discrepancy existing between the maps determined in the two
ways-- the values of ${\cal M}(v_c)$ is very close to $v_\theta$.
The latter result indicate that on looking at the discontinuities
of (the derivative of) the stationary points complexity one
actually find a signature of the phase transition, but this
signature is seen at the potential energy level of the
``inherent'' saddles, not at the instantaneous potential energy
value. In other words, similar to what happens for the slow
dynamics of disordered systems, it seems that are the inherent
saddles properties that determine the phase transitions in mean
field systems.

The paper is organized as follows: in section \ref{sec:thermo} we
present the model and its thermodynamical behavior; in section \ref{topologia}
we study the properties of the stationary points of the potential energy
and calculate their complexity; in section \ref{sellevicine} and \ref{sec:W}
we study the properties of the inherent saddles. Finally, we draw the
conclusions.

\section{The model}
\label{sec:thermo}

The $\phi^4$ mean field model describes $N$ soft spins $\phi_i$ with a mean field
ferromagnetic interaction.
Its thermodynamics, as well as its Langevin and Newton dynamics, have been
studied in the literature, see e.g. Ref.~\cite{schillingDYN}.
The model is defined by the (configurational) Hamiltonian
\bea
\nonumber
&H=\sum_i h(\phi_i) - \frac{J}{2N} \left( \sum_i \phi_i \right)^2 \\
\label{hamiltonian}
&=\sum_i h(\phi_i) - \frac{JNm^2}{2} \\
&h(\phi) = -\frac{\phi^2}{2} + \frac{\phi^4}{4} \nonumber
\eea
where $\phi_i$ are real continuous variables and the magnetization $m$ is
defined as $m=N^{-1} \sum_i \phi_i$.
Its thermodynamics can be exactly solved, as usual in mean field models, in the
thermodynamic limit. Defining
\beq
D\phi_i = d\phi_i \exp \Big( -\beta h(\phi_i) \Big) \ ,
\eeq
the partition function is given by
\bea
&Z_N(T)=\int d\phi_i \ e^{-\beta H(\phi)} = \nonumber
\int D\phi_i \ e^{\beta \frac{J}{2N} \left( \sum_i \phi_i \right)^2 }  \nonumber \\
\label{partizione}
&=N \int dm \ e^{\beta \frac{JNm^2}{2} } \int D\phi_i \delta \left( Nm- \sum_i \phi_i \right)  \\
&=N (2\pi)^{-1} \int dm \ d\hat{m} \ e^{\beta \frac{JNm^2}{2} + i N m \hat{m} }
\int D\phi_i e^{-i \hat{m} \sum_i \phi_i} \nonumber  \\
&=N(2\pi)^{-1} \int dm \ d\hat{m} \ e^{-\beta N f(m,\hat{m})} \nonumber
\eea
having defined
\beq
f(m,\hat{m}) =  -\frac{Jm^2}{2} - i T m \hat{m} - T \log \int d\phi \ e^{-\beta (h(\phi) + i\hat{m} T \phi)}
\eeq
In the thermodynamic limit the free energy is obtained evaluating the integral
in Eq.~(\ref{partizione}) at the saddle point:
\begin{equation}
\label{freeen1}
f(T)= -T \lim_{N\rightarrow \infty} N^{-1} \log Z_N(T) = \max_{m,\hat{m}} f(m,\hat{m})
\end{equation}
The saddle point equations can be written as:
\bea
\label{magnetiz}
&J m = - i T \hat{m} \nonumber \\
&m = \int d\phi \ {\cal P}(\phi) \ \phi = \langle \phi \rangle_{\cal H}
\eea
where we defined the single-particle hamiltonian and the related Gibbs distribution
\bea
\label{singleparticle}
&{\cal H}(\phi) = h(\phi) - J m \phi \nonumber \\
&{\cal Z}=\int d\phi \ e^{-\beta {\cal H}(\phi)} \\
&{\cal P}(\phi) = \frac{ e^{-\beta {\cal H}(\phi)} }{\cal Z} \nonumber
\eea
Having solved the self-consistency equation for the magnetization $m(T)$,
$m = \langle \phi \rangle_{\cal H}$, the free energy is given by Eq.~\ref{freeen1}:
\begin{equation}
f(T) = \frac{J m^2}{2} - T \log {\cal Z}
\end{equation}
As expected, the model undergoes a second order phase transition from a paramagnetic
($m=0$) high-temperature phase to a ferromagnetic ($m \neq 0$) low-temperature phase.
To find the critical temperature one has to expand the self consistency equation
for the magnetization in powers of $m$:
\bea
&m = \frac{\int d\phi \ \phi \ e^{-\beta h(\phi) + \beta J m \phi}}
{\int d\phi \ e^{-\beta h(\phi) + \beta J m \phi}} \\
\nonumber
&= \beta J m \frac{\int d\phi \ \phi^2 \ e^{-\beta h(\phi)}}
{\int d\phi \ e^{-\beta h(\phi)}} + o(m^3) = {\cal A} m + o(m^3)
\eea
The transition temperature $T_c(J)$ is defined by the condition ${\cal A}=1$, that gives:
\begin{equation}
T_c = J \frac{\int d\phi \ \phi^2 \ e^{-\beta_c h(\phi)}}
{\int d\phi \ e^{-\beta_c h(\phi)}}
\end{equation}
The equilibrium potential energy is then given by
\begin{equation}
v(T)=\frac{d (\beta f(T))}{d \beta} = \frac{J m^2}{2} + \int d\phi \ {\cal P}(\phi) \ {\cal H}(\phi)
\end{equation}
We will be interested in the average potential energy at the transition temperature,
that -recalling that $m(T_c)=0$- is given by
\begin{equation}
v_c=v(T_c)=\frac{ \int d\phi \ h(\phi) \ e^{-\beta_c h(\phi)}}
{ \int d\phi \ e^{-\beta_c h(\phi)}}
\end{equation}
The behavior of the critical energy as a function of the coupling
$J$ is reported in Fig.~\ref{criticiJ}.

%%%%%%%%%%%%%%%%%%%%%%%%%%%%%%%%%%%%%%%%%%%%%%%%%%%%%%%%%%%%%%%%%%%%%%%%%%%%%%%%%%%%%%%%%%%%%%%%%%%%%%%%%%%%%%%%%%%%%%%%%%%%%%%%%%%%%%%%%%%%%%%%%%%%%%%%%%%%%%%

\section{Topological properties of the energy surface}
\label{topologia}

\noindent
In this section we will study the properties of the
stationary points (saddles) of the Potential Energy Surface (PES)
of the system, defined by the Hamiltonian (\ref{hamiltonian}). We
will now focus only on the {\it topological} properties of the
saddles, while in section \ref{sellevicine} we will study the
properties of the saddles sampled by the system equilibrated at
temperature $T$. Similar results, although obtained with a
different procedure, have been discussed in Ref.~\cite{russi,tesiBaroni}.

\subsection{Stationary points}

The stationary points $\phi^s$ are defined by the condition
$\nabla H(\phi^s)$$=$$0$, and their order $\nu$ is defined as the
number of negative eigenvalues of the Hessian matrix
${\rm H}_{ij}(\phi^s)$$=$$({\partial^2 H}/{\partial \phi_i
\partial \phi_j})|_{\phi^s}$.  To determine the location of
the stationary points we have to solve the system
\begin{equation}
\frac{\partial H}{\partial \phi_j}=-\phi_j + \phi^3_j - Jm = 0 \hspace{1cm} \forall j
\label{saddef}
\end{equation}
We want to classify the stationary points of $H$ according to
their magnetization $m$ and their energy $v=H(\phi^s)/N$. Thus,
in Eq.~\ref{saddef} we will consider the magnetization $m$ as a
constant; this is exact in the $N\rightarrow \infty$ limit.
Defining $\alpha=J m$ and $\alpha_0 = \frac{2}{3\sqrt{3}}$, the
solutions of the equation $\phi^3 - \phi = \alpha$ are given by
\bea \nonumber
&\phi_+(\alpha)=\frac{2}{\sqrt{3}} \cos \frac{\psi(\alpha)}{3} \\
\label{solu3grado}
&\phi_0(\alpha)=-\frac{2}{\sqrt{3}} \cos \frac{\psi(\alpha)+\pi}{3} \\
\nonumber
&\phi_-(\alpha)=-\frac{2}{\sqrt{3}} \cos \frac{\psi(\alpha)-\pi}{3} \\
\nonumber &\psi(\alpha) = \tan^{-1} \left( \alpha^{-1}
\sqrt{\alpha_0^2-\alpha^2} \right) \eea and are such that
$\phi_0(0)=0$, $\phi_\pm(0)=\pm1$. For $\alpha=\pm \alpha_0$ we
have $\phi_\mp = \phi_0$, while for $|\alpha| > \alpha_0$ two
solutions become complex and only one solution can be accepted. \\
We will now restrict ourselves to the case $|\alpha|<\alpha_0$,
and at the end we will discuss the case $|\alpha| \geq \alpha_0$.
The stationary points of $H$ are obtained by plugging a fraction
$n_+=N_+/N$ of the $\phi_i$ in $\phi_i=\phi_+(\alpha)$, a
fraction $n_0=N_0/N$ in $\phi_0(\alpha)$ and a fraction
$n_-=N_-/N$ in $\phi_-(\alpha)$. Then, the energy $v$ of the
stationary point is given by Eq.~\ref{hamiltonian}:
\begin{equation}
v = \frac{H(\bar{\phi})}{N} = \sum_\xi n_\xi h(\phi_\xi(\alpha)) - \frac{\alpha^2}{2 J}
\end{equation}
where $\xi=(-,0,+)$.
We can now determine the $n_\xi$ imposing the constraints
\bea
\nonumber
&1 = \sum_\xi n_\xi \\
\label{vincoli}
&\alpha = J m = J \sum_\xi n_\xi \phi_\xi(\alpha) \\
\nonumber &v =  \sum_\xi n_\xi h(\phi_\xi(\alpha)) -
\frac{\alpha^2}{2 J} \eea The latter is a linear system that can
be easily solved for any value of $v$, $\alpha$; one must then
impose the additional constraint $n_\xi \in [0,1]$ that restricts
the allowed values of $\alpha$ and $v$. At given energy, we will
have an interval $\alpha \in [\alpha_{min}(v),\alpha_{max}(v)]$
of allowed values of the magnetization. Recalling that a
permutation of the $\phi_i$ does not change neither the
magnetization nor the energy of the stationary point, the number
of stationary points of magnetization $\alpha$ and energy $v$ is
simply given by \bea
&{\cal N}(\alpha,v)=\frac{N!}{N_+!N_0!N_-!} \sim \exp{N\sigma(\alpha,v)} \\
\nonumber &\sigma(\alpha,v) = \lim_{N\rightarrow \infty} N^{-1}
\log {\cal N}(\alpha,v) = -\sum_\xi n_\xi \log n_\xi \eea To compute
the order of the stationary point, we need the expression of the
Hessian matrix. It is given by
\begin{equation}
{\rm H}_{ij}= (3 \phi_i^2 - 1) \delta_{ij} - \frac{J}{N}
\end{equation}
In the thermodynamic limit it becomes diagonal
\begin{equation}
\label{hessdiag}
{\rm H}_{ij}= (3 \phi_i^2 - 1) \delta_{ij}
\end{equation}
One cannot {\it a priori} neglect the contribution of the
off-diagonal terms to the eigenvalues of ${\rm H}$, but one can
prove \cite{tesiBaroni} that their contribution changes the sign
of at most one eigenvalue out of $N$. Neglecting the off-diagonal
contributions, one can easily realize that the number of negative
eigenvalues of ${\rm H}_{ij}$ is given by the number of
$\phi_i=\phi_0(\alpha)$, then $\nu = N n_0(\alpha,v)$.
To summarize, we obtained the following results for $|\alpha|<\alpha_0$: \\
{\it i)} The stationary points are classified according to their
magnetization $m=\alpha/J$ and their potential energy $v$:
from Eq.s~\ref{vincoli} one can determine the fraction $n_\xi(\alpha,v)$ of $\phi_i=\phi_\xi(\alpha)$; \\
{\it ii)} The number of stationary points of magnetization $\alpha$ and energy $v$ is given by $\exp N \sigma(\alpha,v)$,
where $\sigma(\alpha,v)=-\sum_\xi n_\xi(\alpha,v) \log n_\xi(\alpha,v)$; \\
{\it iii)} The stationary points of magnetization $\alpha$ and energy $v$ have order $\nu = N n_0(\alpha,v)$. \\
We will now consider the case $\alpha=\alpha_0$ (the case $\alpha=-\alpha_0$ gives the same results from symmetry
arguments). The equation $\phi^3 - \phi = \alpha_0$ admits only two solutions, namely $\phi_+=2/\sqrt{3}$ and
$\phi_0=-1/\sqrt{3}$. Thus, in this case, we impose only the first two constraints:
\bea
\nonumber
&1 = \sum_\xi n_\xi = n_0 + n_+ \\
\nonumber
&\alpha_0 = J \sum_\xi n_\xi \phi_\xi = - J \frac{n_0}{\sqrt{3}} + J \frac{2 n_+}{\sqrt{3}}
\eea
from which we get $n_0 = \frac{2}{3} \left( 1 - \frac{1}{3J} \right)$ and
$n_+ = \frac{1}{3} \left( 1 + \frac{2}{3J} \right)$. Note that from the condition $n_0,n_+ \in [0,1]$
these stationary points exist only for $J \geq 1/3$. Their energy is given by
\begin{equation}
v_0(J) = - \frac{1}{6} \left( 1 + \frac{5}{9J} \right)
\end{equation}
These stationary points are characterized by an extensive number ($n_0 N$) of zero eigenvalues of the
Hessian matrix associated with $\phi_i=\phi_0$. The remaining eigenvalues are positive as they are associated
with $\phi_i=\phi_+$. \\
Finally, we consider the case $\alpha > \alpha_0$. In this case, there is only one real and positive
solution $\phi_+$  of the equation $\phi^3-\phi=\alpha$, then $\phi_i = \phi_+$ for all $i$ and from
the self-consistency equation $\alpha = J N^{-1} \sum_i \phi_i = J \phi_+$ we get
\begin{equation}
\phi_+^3 - \phi_+ = J \phi_+
\end{equation}
so that $\phi_+ = \sqrt{J+1}$.
Finally, we have to check that $\alpha = J \sqrt{J+1} > \alpha_0$, and this happens only for $J > 1/3$.
Thus, these two points (the latter and the similar one with negative magnetization) exist only for
$J > 1/3$ and represent the absolute (magnetic) minima of the system.
Their energy is given by $v_M = -(1+J)^2/4$.

\subsection{Configurational entropy}

The configurational entropy $\sigma(v)$ of the stationary points
is defined in Eq.~\ref{complexity}. It can be written as

\bea
&\sigma(v) = N^{-1} \log \int_{\alpha_{min}(v)}^{\alpha_{max}(v)} d\alpha \ e^{N \sigma(\alpha,v)} = \\
&{\rm max}_{\alpha \in [\alpha_{min}(v),\alpha_{max}(v)]}
\sigma(\alpha,v) \nonumber \eea
We will neglect the contribution
coming from the absolute minima (their number being non
extensive) and from the points with $\alpha=\alpha_0$ as they
exist only for a particular value of $v$ at which -as we will see-
$\sigma(v)$ displays a singular behavior. Then, for any given
energy $v$ we can find $\bar\alpha(v)$ such that $\partial
\sigma/\partial \alpha = 0$ and
$\sigma(v)=\sigma(\bar\alpha(v))$. Correspondingly, we can define
the average saddle order $\bar{n}(v)=n_0(\bar\alpha,v)$.

\subsection{Euler characteristic}

The Euler characteristic is defined in Eq.~\ref{eulerdef} and can be written as
\begin{equation}
\chi(v)=\int_{-\infty}^v du  \int_{\alpha_m(u)}^{\alpha_M(u)}
d\alpha \ e^{N ( \sigma(\alpha,u) + i \pi n_0(\alpha,u))} \ ,
\end{equation}
recalling that $\nu = N n_0(\alpha,u)$ is the order of the stationary points of magnetization $\alpha$
and energy $u$.
One can attempt to calculate the integral via the saddle point approximation: one has then to find the
stationary points of the function
\begin{equation}
f(\alpha,u)=\sigma(\alpha,u) + i \pi n_0(\alpha,u)
\end{equation}
with respect to the variables $\alpha$ and $u$. Moreover, $\alpha$
and $u$ must be considered as complex variables as the function
$f$ has a non vanishing imaginary part. However, in the model
under discussion, at least at low $v$, the saddle point either do
not exist or is not on a path going from $\alpha_{min}$ to
$\alpha_{max}$ on which $\text{Re}f$ is smaller than its value at
the saddle point. Thus, we expect $\log \chi(v)$ to be
nonextensive at low $v$; in this case the saddle point
approximation is not useful to evaluate $\chi(v)$ and one must
take into account the strong cancellations between addends in
Eq.~\ref{eulerdef}. This point need further investigation and we
will not discuss it here. However, we stress that $\sigma(v)$ is
probably very different from $\chi(v)$ at least at low energy.

\subsection{Summary of the results}

We will now summarize the topological behavior of the model at
different values of $J$. All the results have been obtained
solving numerically the equation $\partial \sigma / \partial
\alpha = 0$ to calculate $\bar\alpha (v)$ and substituting it in
the explicit expressions for all the other interesting quantities.

A first qualitative change in the topology is found at $J_1=1/3$,
while a second at $J_2=1/3$. We will now analyze in some details
the three regions of couplings: weak ($J<J_1)$, intermediate
($J_1<J<J_2)$ and strong ($J>J_2)$.

\subsubsection{Weak coupling}

%%%%%%%%%%%%  TOPO 1 %%%%%%%%%%%%%%%%%%%%%%%%
\begin{figure}
\centering
\includegraphics[width=.45\textwidth,angle=0]{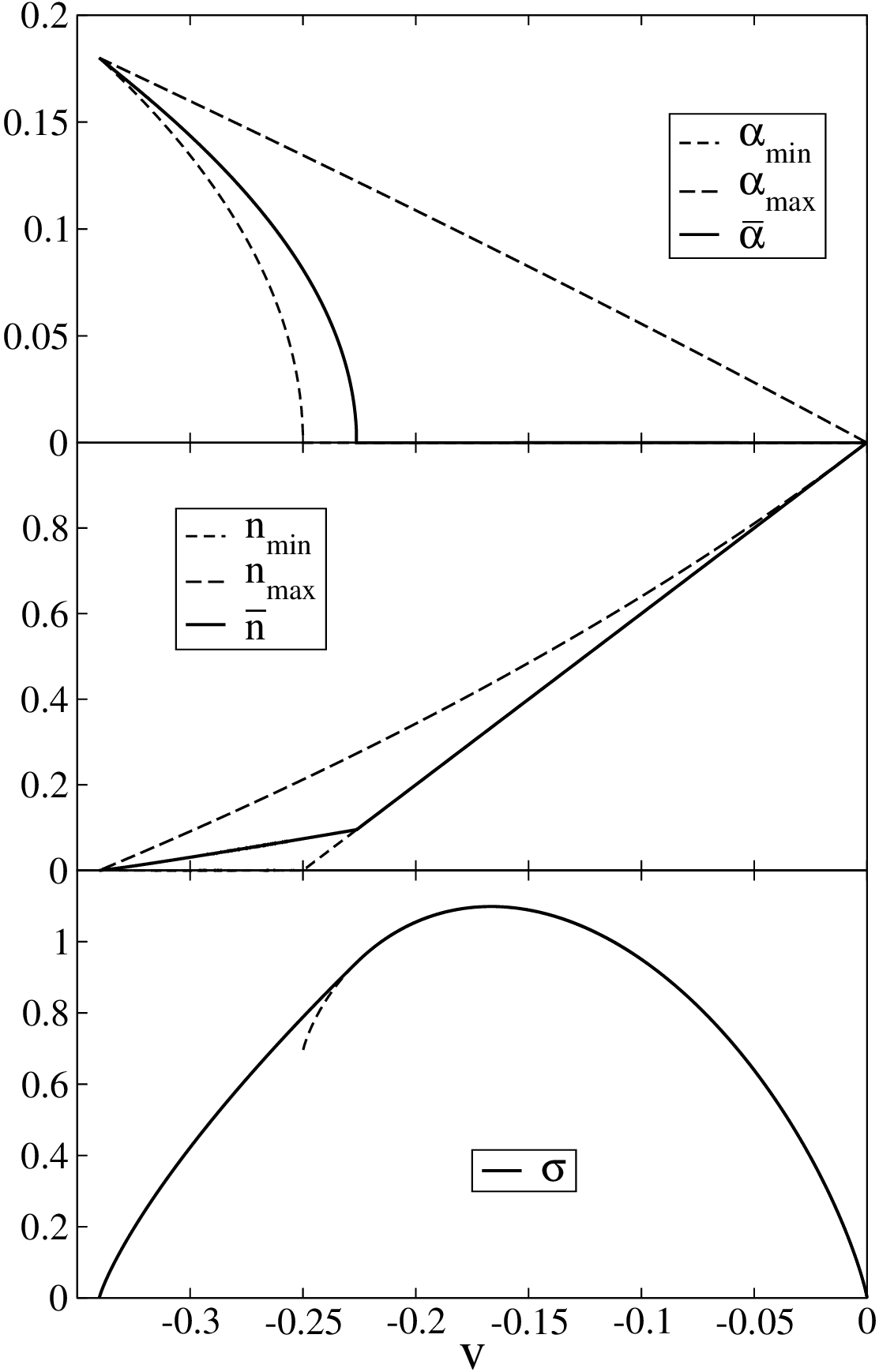}
\caption{Topological properties of the energy surface for $J=1/6$.
Upper panel: maximum and minimum allowed values for the magnetization of the saddles as a function of
their energy (dashed lines) and the value $\bar\alpha(v)$ (full line) that corresponds to the maximum configurational
entropy. Central panel: the maximum and minimum allowed values for the order of the saddles as a function of
their energy (dashed lines) and the value $\bar n(v)$ (full line) that corresponds to the maximum configurational
entropy. Lower panel: total configurational entropy of the saddles as a function of $v$.
For these values of $J$ there is only one singularity $v_\theta$ below which $\bar\alpha \neq 0$.}
\label{topo1}
\end{figure}
%%%%%%%%%%%%%%%%%%%%%%%%%%%%%%%%%%%%%%%%%%%%%%
In Fig.~\ref{topo1} we report the investigated quantities for
$J=1/6 < J_1$. In the upper panel we report, as a function of the
energy $v$, the minimum and maximum allowed values of $\alpha$
(dashed lines), together with the value $\bar\alpha(v)$ determined
by the maximization of $\sigma(\alpha,v)$ (full line). Above
$v=-1/4$ it results $\alpha_{min} = 0$, while below $v=-1/4$ the
paramagnetic ($\alpha=0$) stationary points disappear and
$\alpha_{min} > 0$. In this $J<1/3$ region we have $\alpha_{max} <
\alpha_0$ for any $v$. In the central panel we report the saddle
order as a function of the energy. Above $v=-1/4$ there are no
minima ($n_{min}>0$) while below $v=-1/4$ minima and saddles
coexist. The absolute minima are at $v_M \sim -0.34$, where
$n_{max} \rightarrow 0$. Thus, there exist saddles of order $\nu >
0$ arbitrary close (in energy) to the absolute minima. In the
lower panel we report the configurational entropy as a function of
$v$. From the upper panel we see that there exist a value
$v_\theta>-1/4$ above which $\bar\alpha(v)=0$ while for
$v<v_\theta$ we have $\bar\alpha(v)>0$. At the same point the
configurational entropy displays (obviously) a singularity: the
curve $\sigma(0,v)$ is reported as a dashed line. Note that even
if for $v_M < v < -1/4$ there exist stationary points with
fractional order $\nu/N=0$, we have $\bar{n}(v)>0$.

\subsubsection{Intermediate coupling}

%%%%%%%%%%%%  TOPO 2 %%%%%%%%%%%%%%%%%%%%%%%%
\begin{figure}
\centering
\includegraphics[width=.45\textwidth,angle=0]{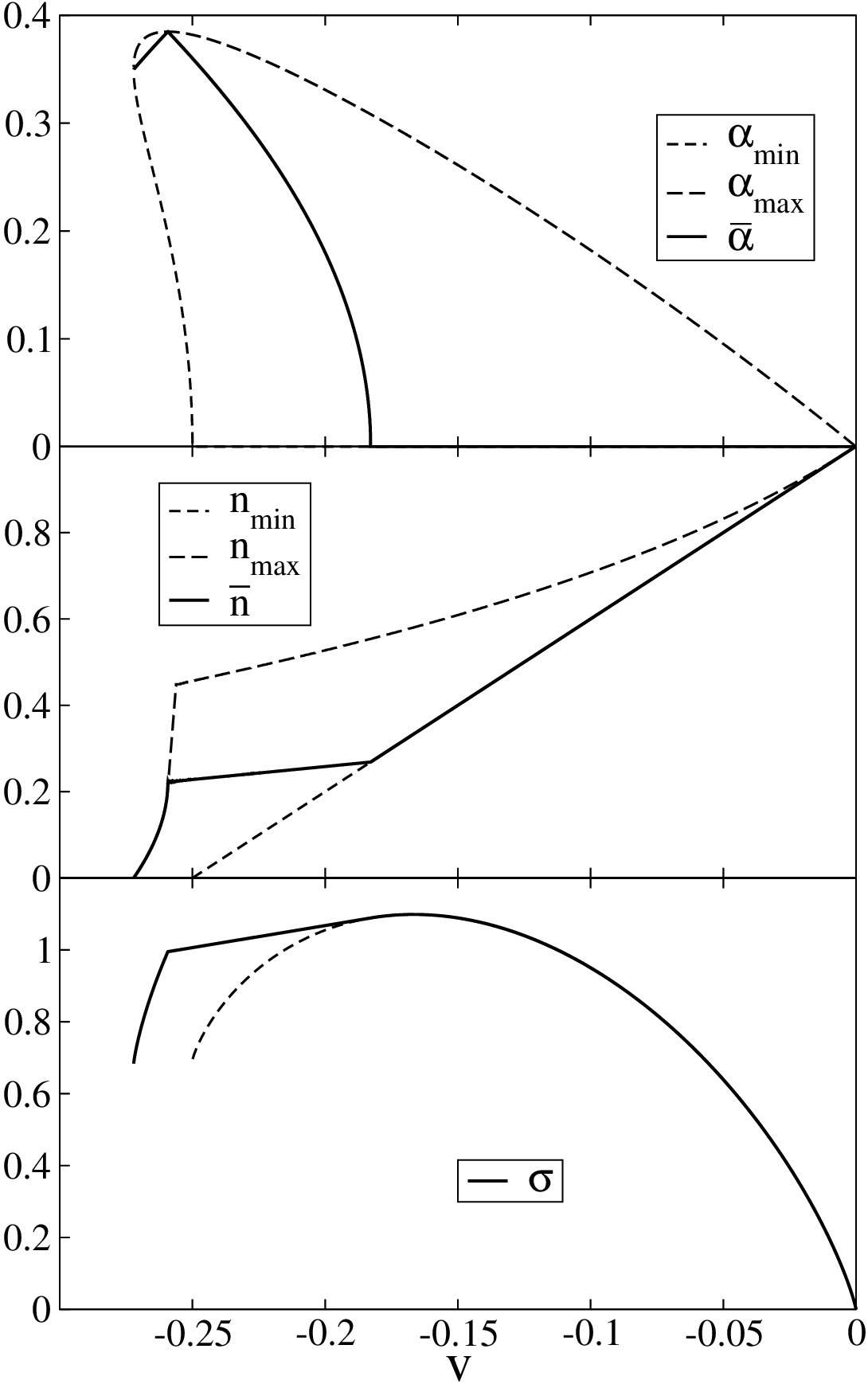}
\caption{Topological properties of the energy surface for $J=1$. The plots are the same as in Fig.~\ref{topo1}.
In this region a second singularity $v_0$ appears where $\bar\alpha=\alpha_0$.
Below $v_0$ $\bar\alpha$
decreases again until $v$ reaches its lowest possible value. In this region, the absolute minima are far
below the minimum energy of the saddles and are not reported in the figure (see text).}
\label{topo2}
\end{figure}
%%%%%%%%%%%%%%%%%%%%%%%%%%%%%%%%%%%%%%%%%%%%%%
%%%%%%%%%%%%  TOPO 3 %%%%%%%%%%%%%%%%%%%%%%%%
\begin{figure}
\centering
\includegraphics[width=.45\textwidth,angle=0]{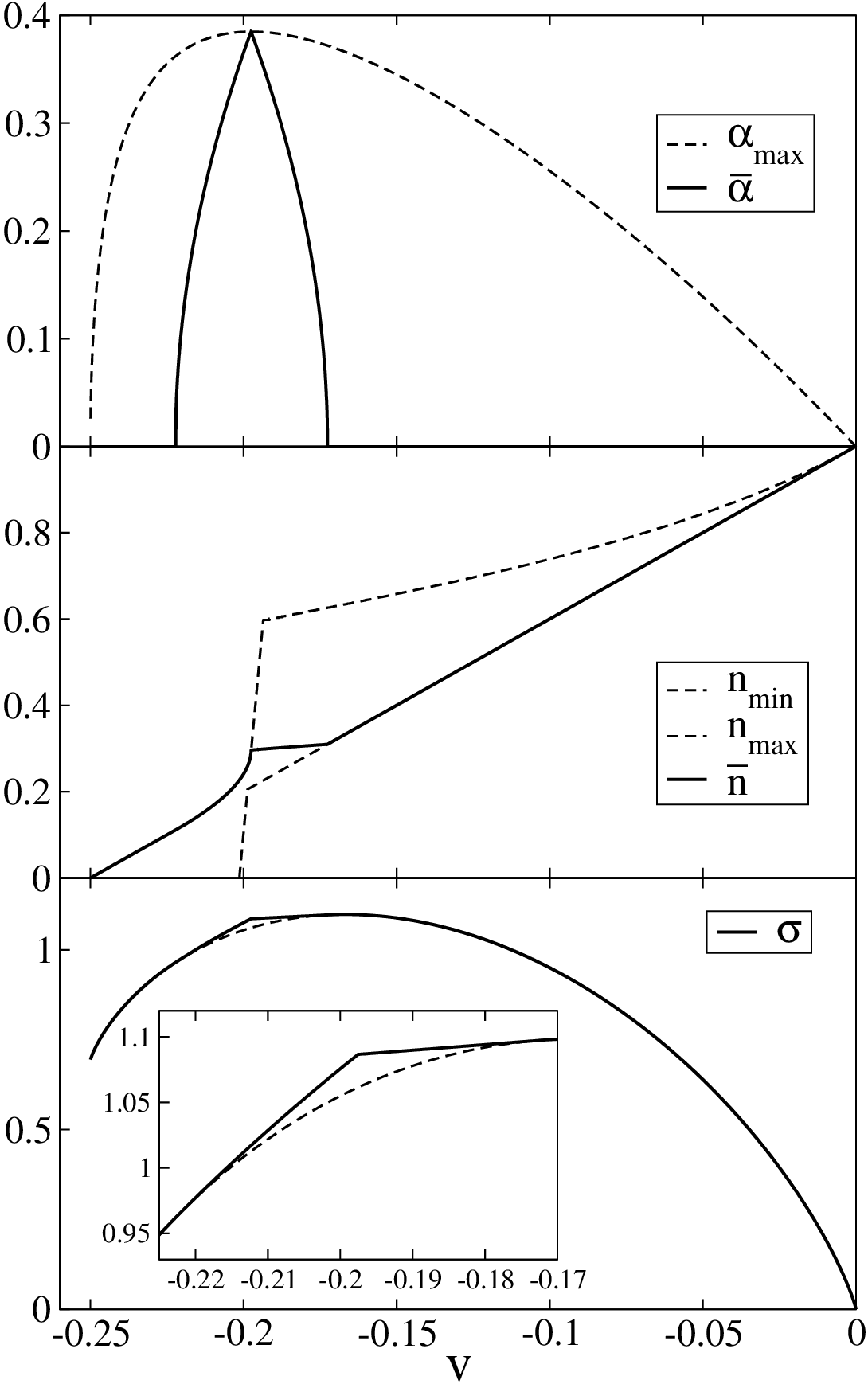}
\caption{Topological properties of the energy surface for $J=3$. The plots are the same as in Fig.~\ref{topo1}.
As in the previous figures there are two singularities of $\sigma(v)$ at $v_\theta$ and $v_0$.
Moreover, in this region a third singularity $v_2$ appears below which again $\bar\alpha=0$.
As in Fig.~\ref{topo2}, the absolute minima are not reported in the figure.
In the inset of the lower panel the region of the three singularities is magnified: one sees that $\sigma(v)$
is different from $\sigma(\alpha=0,v)$ only the interval $v \in [v_2,v_\theta]$.}
\label{topo3}
\end{figure}
%%%%%%%%%%%%%%%%%%%%%%%%%%%%%%%%%%%%%%%%%%%%%%
In Fig.~\ref{topo2} the same quantities of Fig.~\ref{topo1} are
reported for $J_1 \leq J \leq J_2 =2$ (namely, $J=1$). Again, we
have a singularity at $v_\theta > -1/4$ where $\bar\alpha$ become
different from 0. Moreover, in this region, the points with
$\alpha=\alpha_0$ appear: as we can see from the upper panel, both
$\alpha_{max}$ and $\bar\alpha$ move toward $\alpha_0$ for $v
\rightarrow v_0$. At $v=v_0$, we find $\bar\alpha=\alpha_0$, then
for $v<v_0$ $\bar\alpha$ starts to decrease. The configurational
entropy (lower panel) shows two singular points, the first at
$v_\theta$ and the second at $v_0$. In the central panel the order
of the saddles is reported. In this case, the minima are located
at $v_M=-1$, well separated from the lowest order saddles. Then,
in this case, a gap between the absolute minima and the lowest
order saddles opens and $\bar{n}(v)$ goes to zero at a value $v >
v_M$.

\subsubsection{Strong coupling}

At $J_2 = 2$ a third singularity $v_2<v_0<v_\theta$ appears, below
which $\bar\alpha=0$ and again the paramagnetic saddles dominate.
In Fig.~\ref{topo3} we report the results for $J=3 > J_2$. We note
that in this region we always have $\alpha_{min}=0$, while
$\bar\alpha$ is zero for $v>v_\theta$, increases toward $\alpha_0$
for $v_0<v<v_\theta$ and then decreases again and reaches zero at
$v=v_2$, as previously discussed. The configurational entropy then
follows the $\alpha=0$ curve apart from the interval
$[v_2,v_\theta]$ in which it shows the additional singularity at
$v_0$. In the inset of the lower panel we show the behavior of
$\sigma(v)$ in the interval $[v_2,v_\theta]$. Again the absolute
minima are at very low energy ($v_M=-4$) and are well separated
from the lowest order saddles.

\subsection{Discussion}

As we discussed in the introduction, it has been conjectured and
verified in many different models \cite{pettinisumma} that
topological singularities could be related to thermodynamic
singularities (phase transitions) or dynamic singularities (glass
transitions). We showed that the model has a very complex
topological behavior. In particular, for $J<J_1$ there is only one
singularity at $v=v_\theta$ below which the saddles are
characterized by a ``spontaneous magnetization''; for $J_1<J<J_2$
another singularity appears at $v=v_0<v_\theta$; the latter is due
to the presence, at $v=v_0$, of points with magnetization
$\alpha=\alpha_0$, characterized by a large number of zero
eigenvalues of the Hessian matrix. For $J>J_2$, a third singular
point $v_2$, below which the paramagnetic saddles again dominate,
appears. However, for our discussion only $v_\theta$ will be
relevant, as it represents the energy below which the saddles with
$\alpha \neq 0$ become dominant, and hence could be expected to be
related to the thermodynamical phase transition. If this is the
case, one could expect the thermodynamical critical energy $v_c$
to be close to $v_\theta$.

In Fig.~\ref{criticiJ} we report $v_\theta$ (full line) as a
function of the coupling $J$ toghether with the thermodynamical
critical energy $v_c$ (dot-dashed line). One immediately notices
that $v_c$ is far above $v_\theta$, at variance to what is found
in the previously investigated mean field models
\cite{PettiniXY,noieulero}; moreover, at high $J$ one has $v_c>0$
while there are no stationary points of the Hamiltonian at
positive energy, as already recognized in Ref.~\cite{tesiBaroni}.
From this argument and from Fig.~\ref{criticiJ} one concludes
that, to relate the phase transition to changes in the topology of
the PES one has to generalize in a suitable way the relation $v_c
\sim v_\theta$ found in \cite{PettiniXY,noieulero}. This will be
the aim of the next section.

%%%%%%%%%%%%%%%%%%%%%%%%%%%%%%%%%%%%%%%%%%%%%%%%%%%%%%%%%%%%%%%%%%%%%%%%%%%%%%%%%%%%%%%%%%%%%%%%%%%%%%%%%%%%%%%%%%%%%%%%%%%%%%%%%%%%%%%%%%%%%%%%%%%%%%%%%%%%%%%%%%%%%%%%%%%%%%%%%%%%%%%%%%%%%%%%%%%%%%%%%%%%%%%%%%%

\section{Inherent saddles}
\label{sellevicine}

\noindent Recent works established that, in order to describe the
equilibrium dynamics at a given temperature $T$, it is sufficient
to know the properties of some of the stationary points, that have
often been called ``inherent saddles''
\cite{noi_sad,cav_sad,cav-pspin,parisi_boson}. To locate these
particular stationary points, two main strategies have been
adopted in the past: {\it (1)} partitioning the phase space in
``basins of attraction'' of stationary points via an appropriate
function that has a local minimum on each stationary point; {\it
(2)} defining in a proper way a ``distance'' in phase space and,
given an equilibrium configuration, looking at the stationary
point that has minimum distance from this configuration. It has
been shown in \cite{noiptrig} that, at least in a simple mean
field model, these two definition give exactly the same result. In
this section, we will discuss the properties of the inherent
saddles using definition {\it 2)}, which is more suitable for
analitical calculations, and later compare the results with the
one obtained using definition~{\it 1)}.

To calculate the average energy and magnetization of the closest saddles to
equilibrium configurations,
we will make use of the method introduced
in \cite{cav-pspin}. We compute the quantity
\begin{eqnarray}
\nonumber &\Sigma(T;v_s,d) =\frac{1}{N} \int d\phi_i
\frac{e^{-\beta
H (\phi)}}{Z(T)} \log \big [ \int d \psi_i \ \delta(H(\psi)-Nv_s) \\
&\delta(\partial_i H(\psi)) \ |\det {\rm H}(\psi)| \ \delta \left(
d^2 - d^2(\phi,\psi) \right) \big ] \nonumber
\end{eqnarray}

where ${\rm H}_{ij}=\partial_i \partial_j H$ is the Hessian matrix
and $d(\phi,\psi)$ is a distance function between the two
configurations $\phi_i$ and $\psi_i$. The argument of the
logarithm is the number of stationary points of energy $v_s$ and
distance $d$ from the reference configuration $\phi$ (see
reference~\cite{cav-pspin,noiptrig} for a detailed discussion).
Then the logarithm of this number (divided by $N$) is averaged
over the equilibrium distribution at temperature $T$ of the
reference configuration. To find the closest saddles to
equilibrium configurations -at given temperature $T$- we must find
the minimum $d$ such that $\Sigma(T;v_s,d) \geq 0$ (otherwise the
number of saddles at distance $d$ is zero). The condition
$\Sigma(T;v_s,d) \geq 0$ will define a domain ${\cal D}_+$ in the
$(v_s,d)$ plane. We have then to find the minimum $d(T)$ of $d$ in
${\cal D}_+$. Usually, this will correspont to a single value of
$v_s$, that will be called $v_s(T)$ and represents the energy of
the closest saddles. Note also that the point $(v_s(T),d(T))$ will
be on the border of the domain ${\cal D}_+$ that is defined by
the condition $\Sigma(T;v_s,d) \geq 0$, thus $\Sigma(T;v_s(T),d(T)) = 0$ \cite{cav-pspin,noiptrig}. \\
In our model the distance function can be defined as
\begin{equation}
 d^2(\varphi,\psi)=\frac{1}{N} \sum_i (\phi_i-\psi_i)^2
\end{equation}
%%%%%%%%%%%%  TEMPERATURA weak %%%%%%%%%%%%%%%%%%%%%%%%
\begin{figure}
\centering
\includegraphics[width=.47\textwidth,angle=0]{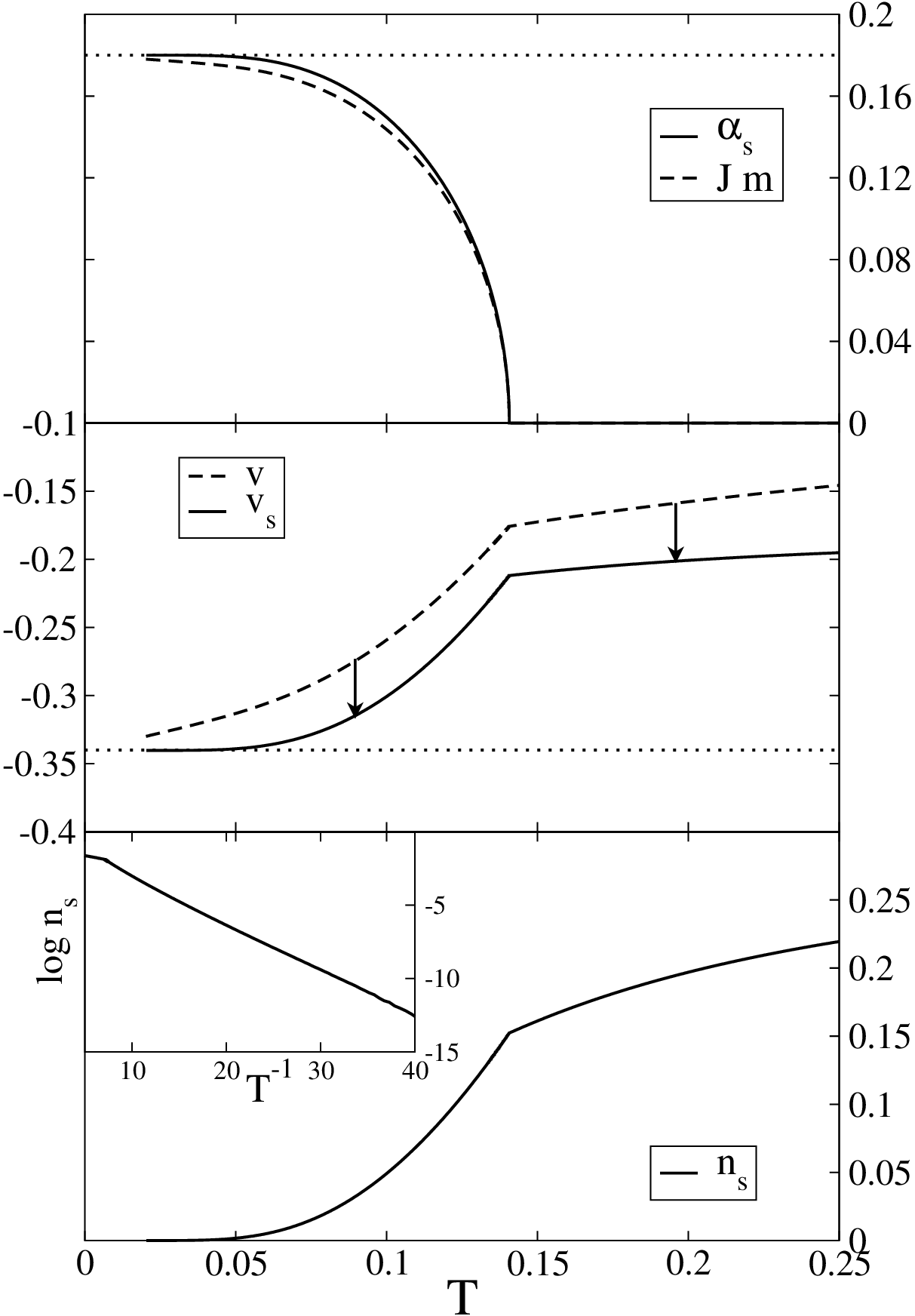}
\caption{Properties of the inherent saddles for $J=1/6$ at different temperatures $T$.
Upper panel: magnetization $\alpha_s(T)$ of the closest saddles and magnetization $J m(T)$ of the equilibrium
configurations.
Central panel: thermodynamical energy $v(T)$ and energy $v_s(T)$ of the closest saddles. The arrows graphically
show the mapping ${\cal M}$ of the istantaneous energy into the inherent saddles energy.
Lower panel: saddle order $n_s(T)$. In the inset, $\log n_s$ is reported as a function of $T^{-1}$ to enhance the
low temperature Arrhenius behavior, $\log n_s = -\Delta/T$.
}
\label{selleT1}
\end{figure}
%%%%%%%%%%%%%%%%%%%%%%%%%%%%%%%%%%%%%%%%%%%%%%
The direct calculation of $\Sigma(T;v_s,d)$ is reported in
appendix \ref{app_distance}. There we show that the energy,
distance and magnetization of the closest saddles as a function of
the temperature are given by the solution of the following
equations: \bea \nonumber
&\alpha = J \int d\phi \ {\cal P}(\phi) \ \tilde\phi(\phi,\alpha) = f(\alpha,T) \\
&d^2(T) = \int d\phi \ {\cal P}(\phi) \ [\tilde\phi(\phi,\alpha) - \phi ]^2
\label{lepiuvicine} \\
\nonumber &v_s(T) = \frac{\alpha^2}{2J} + \int d\phi \ {\cal
P}(\phi) \ {\cal H}(\tilde\phi(\phi,\alpha)) \eea where the
function $\tilde\phi(\phi,\alpha)$ is equal to the
$\phi_\xi(\alpha)$ such that $(\phi_\xi(\alpha)-\phi)^2$ is
minimum, and ${\cal P}$ has been defined in
equation~(\ref{singleparticle}). The first equation has to be
interpreted as a self-consistency equation for $\alpha$ which
solution is the magnetization $\alpha_s(T)$ of the inherent
saddles. Substituting $\alpha_s(T)$ in the second and third
equation one gets the average distance $d(T)$ between equilibrium
configurations and inherent saddles and the average energy
$v_s(T)$ of the inherent saddles. Finally, substituting
$\alpha_s(T)$ and $v_s(T)$ in the expression for the number of
saddles and for their order derived in section~\ref{topologia} we
get the configurational entropy $\sigma(T)$ and the order $n_s(T)$
of the inherent saddles:

\bea
&\sigma(T)=\Sigma(\alpha_s(T),v_s(T)) \\
&n_s(T)=n_0(\alpha_s(T),v_s(T)) \nonumber
\eea

Note that $\sigma(T)$ has not to be confused with
$\sigma(T;v_s(T),d(T))=0$. In fact, the latter is the number of
saddles of energy $v_s$ subject to the additional constraint of
having distance $d$ from the equilibrium configurations, while the
first is simply the number of saddles of energy $v_s$ and
magnetization $\alpha_s$.

\subsection{Properties of the inherent saddles}

We will now discuss the properties of the inherent saddles in the
weak and strong coupling regimes. We numerically solve the first
of Eqs.~\ref{lepiuvicine} to get $\alpha_s(T)$, and from the other
two we get all the quantities of interest.

\subsubsection{Weak coupling}

%%%%%%%%%%%%  TEMPERATURA strong %%%%%%%%%%%%%%%%%%%%%%%%
\begin{figure}
\centering
\includegraphics[width=.45\textwidth,angle=0]{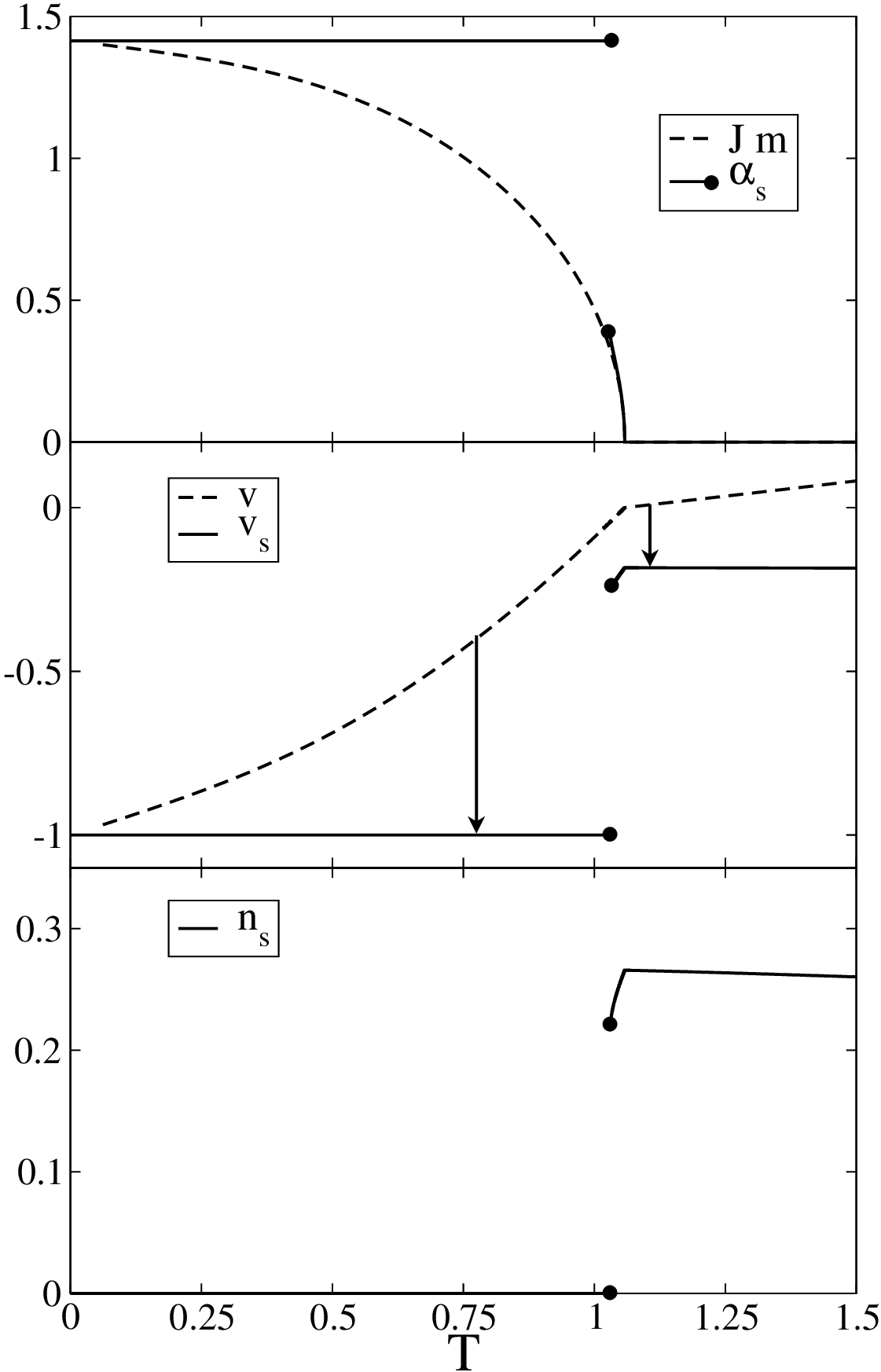}
\caption{Properties of the inherent saddles for $J=1$ at different temperatures $T$.
The plots are the same as in Fig.~\ref{selleT1}. In the strong coupling regime the system jumps
discontinuously in the minima at temperature $T^*$ (marked by black dots in the figure).
}
\label{selleTstr}
\end{figure}
%%%%%%%%%%%%%%%%%%%%%%%%%%%%%%%%%%%%%%%%%%%%%%

The behavior of the investigated quantities as a function of the
temperature for $J=1/6$ is reported in Fig.~\ref{selleT1}. In the
upper panel, the magnetization $\alpha_s(T)=Jm_s(T)$ of the
closest saddles is reported together with the thermodynamic
magnetization $J m(T)$. We notice that $m_s(T) \sim m(T)$: thus,
the system visits saddles that have a magnetization very similar
to the equilibrium one. At low temperature, the system stays very
close to the absolute minima (whose magnetization is reported as a
dotted line) even if it reaches them only at $T=0$. In the central
panel, we report the energy $v_s(T)$ of the inherent saddles
(dotted line) and the equilibrium energy $v(T)$ (full line). At
$T=T_c$, both $v(T)$ and $v_s(T)$ show a singular behavior. We can
observe that, in the present model, the saddle energy at $T_c$ is
smaller than the equilibrium energy, i.e. $v_s(T_c) = {\cal
M}(v_c) < v_c$. This findings is at variance whit the $XY$ and
$k$-trigonometric models where one finds ${\cal M}(v_c)=v_c$
\cite{noiptrig}. We observe that the value of $v_s(T_c)$, for
$J=1/6$, turns out to be $v_s(T_c)$=XXX, very close to
$v_\theta$=XXX. At low temperature $v_s(T)$ is very close to the
energy of the absolute minima. Finally, in the lower panel, we
report the saddle index $n_s(T)$. From the inset we see that, for
$T\sim 0$, $n_s(T)$ has an Arrhenius behavior, $n_s(T) \sim
\exp(-\Delta/T)$ \cite{noiptrig}.

\subsubsection{Intermediate coupling}

%%%%%%%%%%%%  MAPPA %%%%%%%%%%%%%%%%%%%%%%%%
\begin{figure}
\centering
\includegraphics[width=.45\textwidth,angle=0]{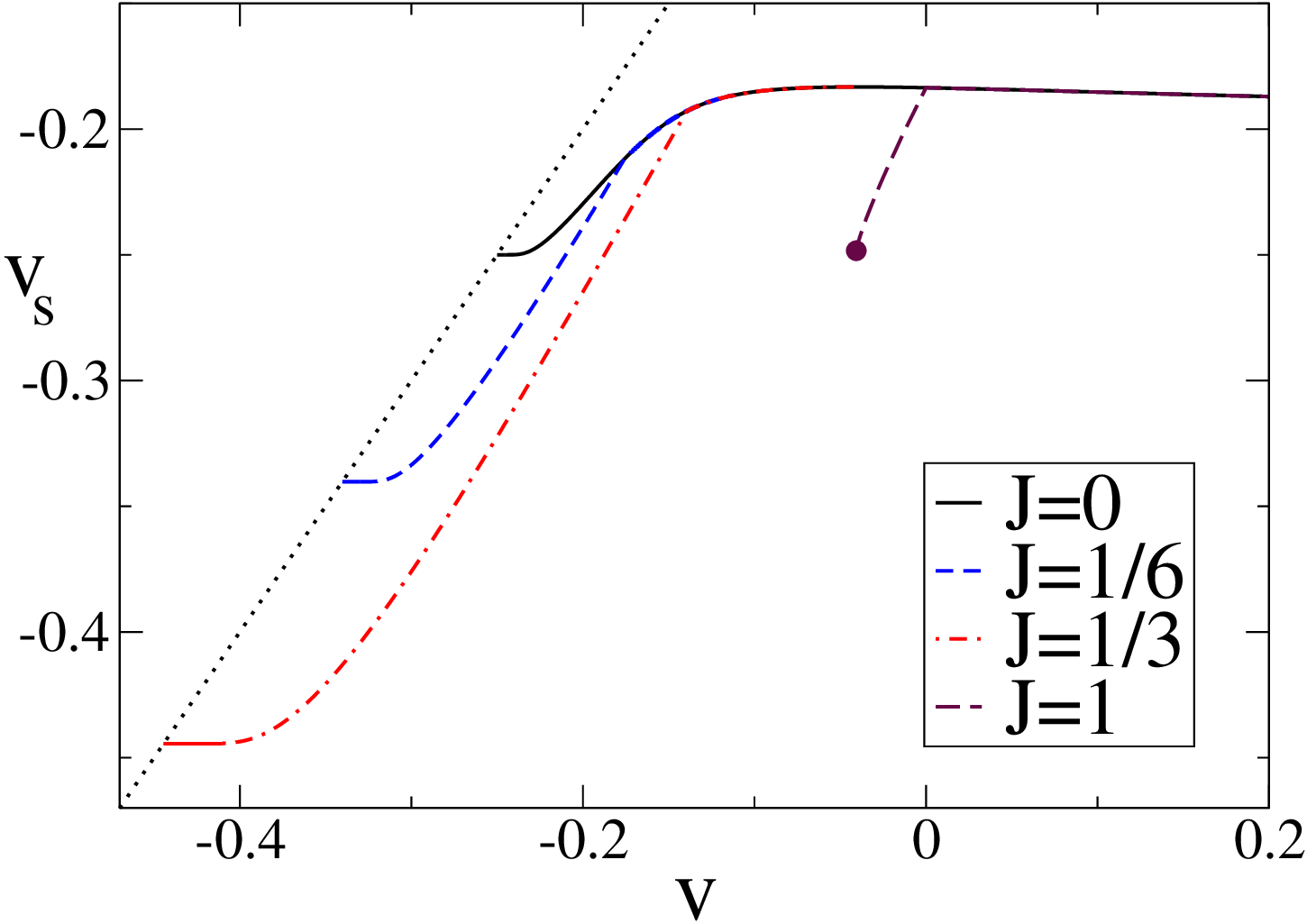}
\caption{
The function ${\cal M}(v)$ that associate to the thermodynamic energy $v$ the
corresponding inherent saddle energy $v_s$ for different values of $J$.
The system follow the $J=0$ curve in the paramagnetic phase while below
$T_c$ the inherent saddles have lower energy with respect to $J=0$.
At $T=0$ the system is in the minima and $v_s={\cal M}(v)=v$ (dotted line).
For $J > 1/3$ the system jumps discontinuously into the minima at a certain
energy $v^*=v(T^*)$ (full dot in the figure).
}
\label{mappaGCR}
\end{figure}
%%%%%%%%%%%%%%%%%%%%%%%%%%%%%%%%%%%%%%%%%%%%%%

In the intermediate and strong coupling regime ($J>1/3$) the
topology of the PES is very complicated, as we showed in the
previous section (see Figs.~\ref{topo2} and \ref{topo3}). In
particular, in this region the minima are separated from the lower
energy saddles by a gap and two (or three) topological
singularities appear. In Fig.~\ref{selleTstr} we report as an
example the behavior of the magnetization, energy and order of the
inherent saddles as a function of the temperature for $J=1$. We
see that as in the weak coupling regime the system samples
non-magnetized saddles above $T_c$ while below $T_c$ one has
$\alpha_s(T) \neq 0$. However, at a given temperature $T^*$ the
system jumps discontinuously into the minima: below $T^*$ the
saddle order is exactly $0$, the energy is $v_s(T)=v_M=-(1+J)^2/4$
and the magnetization is $\alpha_s=J \sqrt{1+J}$. On increasing
$J$, $T^*$ moves toward $T_c$. Note that there is no qualitative
difference between the intermediate ($J<2$) and strong ($J>2$)
coupling as the low energy saddles that are sligthly below the gap
are never visited by the system.

\subsection{Mapping the instantaneous energy into the inherent saddles energy}

As we discussed in the previous section, the equilibrated system
at temperature $T$ is close to saddles that have energy $v_s(T) <
v(T)$, where $v(T)$ is the thermodynamical energy. We can
construct the function ${\cal M}$ that maps the instantaneous
energy $v$ into the inherent saddles energy $v_s={\cal M}(v)$
using the temperature as a parameter. The function ${\cal M}$ is
reported in Fig.~\ref{mappaGCR} as a function of $v$ for selected
$J$ values. To check whether the energy of the inherent saddles at
$T_c$ is close to the singularity $v_\theta$, we need to compute
$v_s(T_c)={\cal M}(v_c)$.
%%%%%%%%%%%%  RIASSUNTO PUNTI CRITICI %%%%%%%%%%%%%%%%%%%%%%%%
\begin{figure}
\centering
\includegraphics[width=.45\textwidth,angle=0]{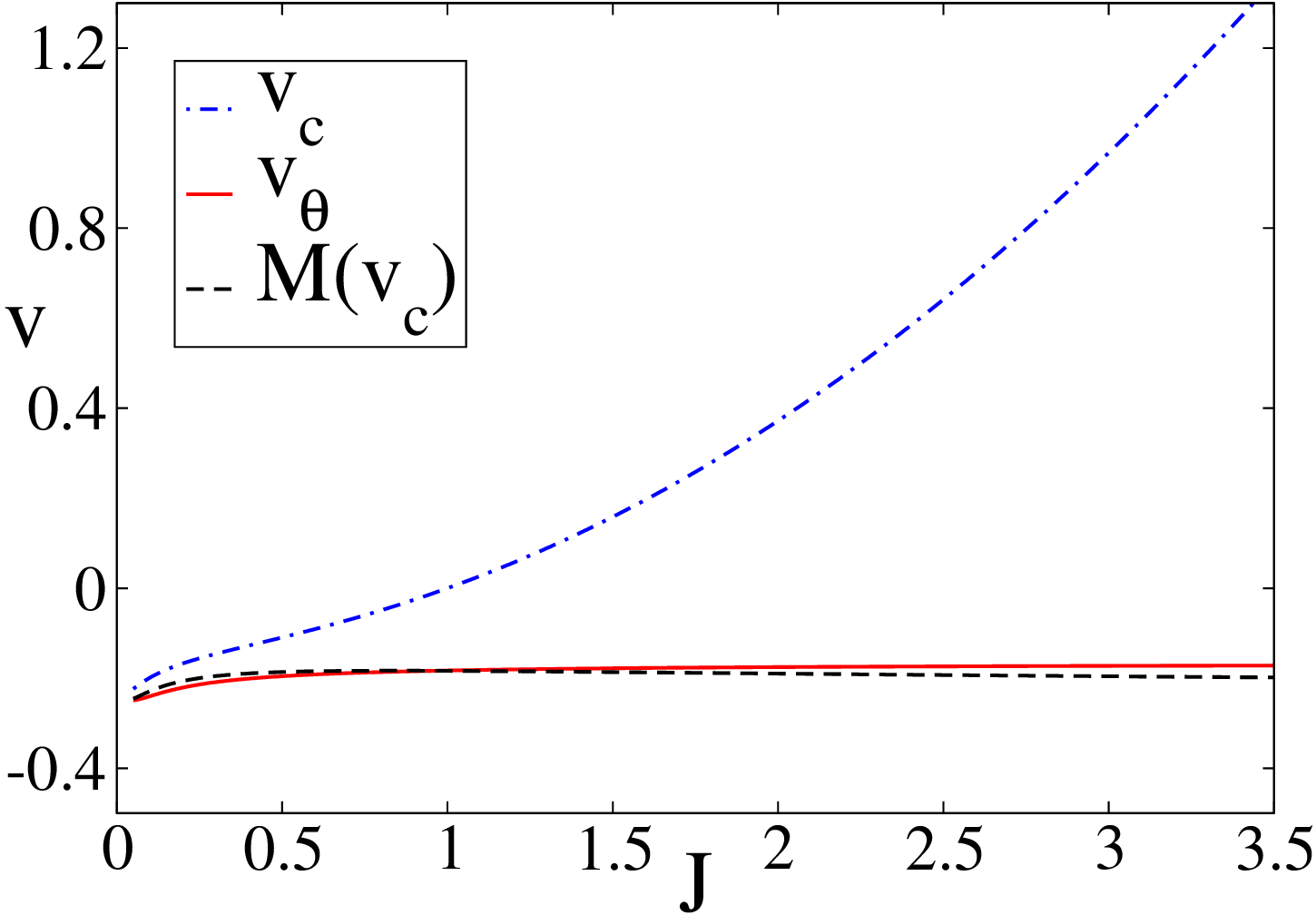}
\caption{
The thermodynamical transition point $v_c=v(T_c)$ (dot-dashed line),
the singularity of $\sigma(v)$, $v_\theta$ (full line),
and the energy of the inherent saddles at $T_c$, $v_s(T_c)={\cal M}(v_c)$ (dashed line),
as functions of the coupling $J$.
}
\label{criticiJ}
\end{figure}
%%%%%%%%%%%%%%%%%%%%%%%%%%%%%%%%%%%%%%%%%%%%%%
We can obtain an explicit expression for $v_s(T_c)$ recalling
that, for $T \geq T_c(J)$, we have $m(T)=0$, and ${\cal
H}(\phi)=h(\phi)$. It is easy to see from Eq.~\ref{lepiuvicine}
that $m=0$ implies $\alpha_s(T)=0$. Thus, in the paramagnetic
phase the inherent saddles are always paramagnetic and their
energy is given by the simple expression
\begin{equation}
\label{vsTcdist}
v_s(T\geq T_c) = -\frac{1}{4} \left(1 - \frac{\int_{-1/2}^{1/2} d\phi \ e^{-\beta h(\phi)}}
{\int_{-\infty}^{\infty} d\phi \ e^{-\beta h(\phi)}} \right)
\end{equation}
The energy of the saddles sampled at $T_c$ is simply given by the latter expression calculated
in $T=T_c$.
The energy of the inherent saddles at the critical temperature
as a function of $J$ is reported in Fig.~\ref{criticiJ}.

\subsection{Discussion}

As we showed in section \ref{topologia}, for the model
investigated here, the energy at which the configurational entropy
of the saddles shows a singularity ($v_\theta$) is different from
the energy at which the thermodynamical transition takes place
($v_c$) (see Fig.~\ref{criticiJ}). Recent studies of the dynamics
of glassy systems~\cite{noi_sad,cav_sad,cav-pspin} demonstrated
that the equilibrium properties at temperature $T$ (and energy
$v$) are related to the topological properties of the PES at
energy $v_s={\cal M}(v)$, i.e. the energy of the inherent saddles.
If this is the case, one should expect the phase transition to
happen when the energy of the inherent saddles (and not the
thermodynamical energy) is close to the singularity $v_\theta$ of
$\sigma(v)$. Indeed, as we can see from Fig.~\ref{criticiJ}, the
relation ${\cal M}(v_c) \sim v_\theta$ holds, even if not exactly.
This result, the main finding of the present work, generalizes the
relation $v_c \sim v_\theta$, discussed in
Ref.~\cite{PettiniXY,noieulero,pettinisumma,TeoremaPettini} for
cases where ${\cal M}(v_c) \sim v_c$, to the present case where
${\cal M}(v_c) \neq v_c$.

%%%%%%%%%%%%  ORDINE ENERGIA %%%%%%%%%%%%%%%%%%%%%%%%
\begin{figure}
\centering
\includegraphics[width=.45\textwidth,angle=0]{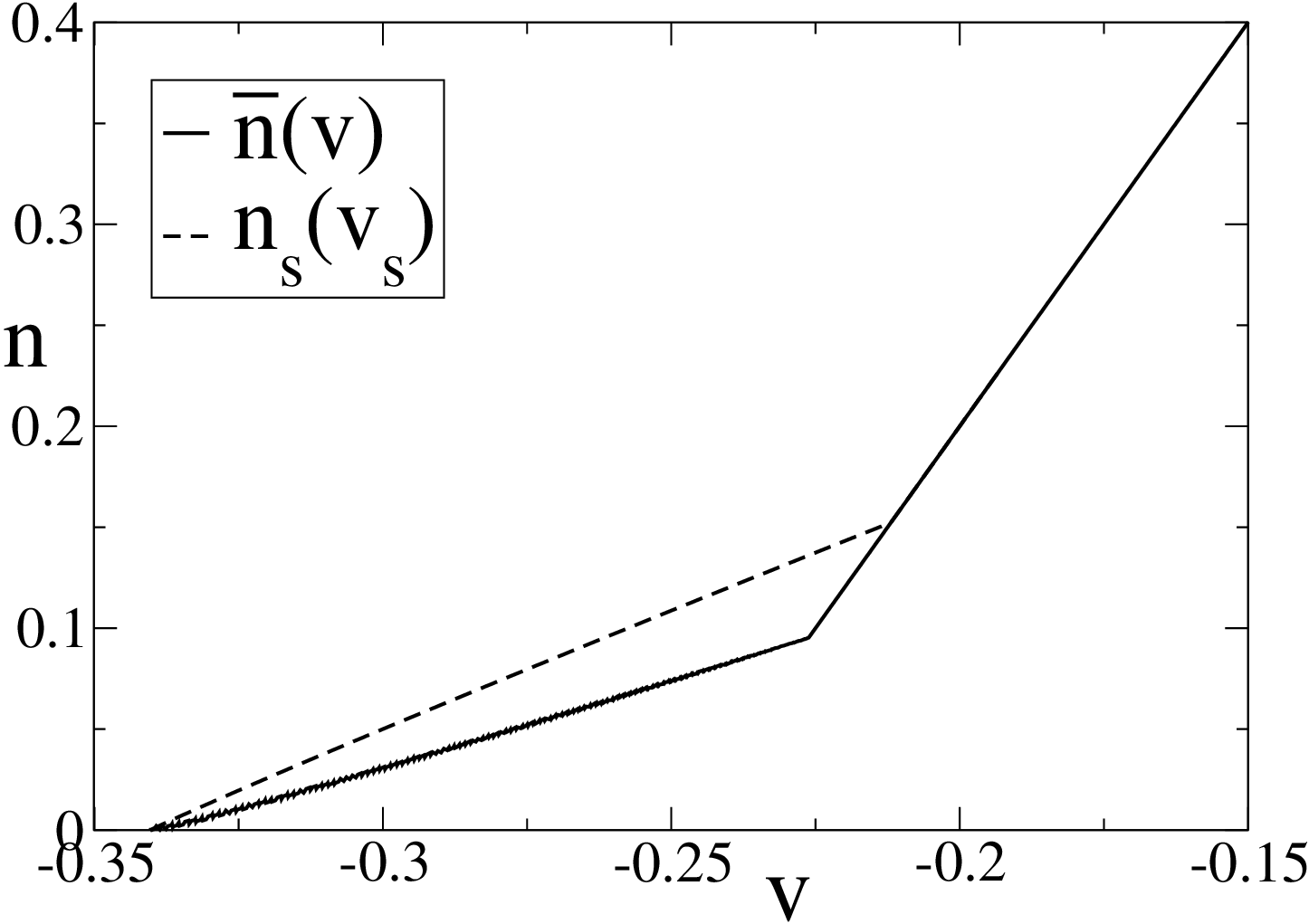}
\caption{ Comparison between the order of the closest saddles
(full line) and the order of the dominant ones (dashed line, see
text) for $J=1/6$. $n_s(T)$ is reported as a function of $v_s(T)$
(see Fig.~\ref{selleT1}) parametrically in $T$ while $\bar n(v)$
is the same as in Fig.~\ref{topo1}. } \label{selleT2}
\end{figure}
%%%%%%%%%%%%%%%%%%%%%%%%%%%%%%%%%%%%%%%%%%%%%%

To better understand the origin of the small difference between
${\cal M}(v_c)$ and $v_\theta$, in Fig.~\ref{selleT2} we report
(for $J=1/6$) the saddle order as a function of the saddle energy
for {\it i)} the saddles that dominate in the configurational
entropy and {\it ii)} the inherent saddles. As we clearly see from
Fig.~\ref{selleT2}, the system is not always close to the saddles
of order $\bar n$ that dominate in the configurational entropy (in
the following, {\it dominant saddles}); indeed, below $T_c$ the
system start to sample saddles that are subdominant in the
configurational entropy. However, one could expect the system to
be always visiting the dominant saddles at energy $v_s(T)$, as the
number of the dominant saddles is exponentially bigger than the
number of all the other saddles. This discrepancy can be a
consequence of an incorrect definition of the ``basin of
attraction'' of a saddle, i.e. of an incorrect mapping between
equilibrium configuration and inherent saddles. In next section,
we will discuss a different definition of basin of attraction.

Finally, we observe that the physical interpretation of the
discontinuous jump into the absolute minima that occurs for
$J>1/3$ is possibly related to the dynamical behavior of the
system; the clarification of this point requires then the
investigation of the dynamics of the system below $T_c$.
Unfortunately, in Ref.~\cite{schillingDYN} the Langevin dynamics
of the system was studied only above $T_c$; the comparison of our
results with the dynamical behavior of the system requires the
extension of the calculations in Ref.~\cite{schillingDYN} to the
ferromagnetic phase.

\section{Depence on the definition of basin of attraction of a saddle}
\label{sec:W}

%%%%%%%%%%%%  CRITICI RIASSUNTO + W %%%%%%%%%%%%%%%%%%%%%%%%
\begin{figure}
\centering
\includegraphics[width=.45\textwidth,angle=0]{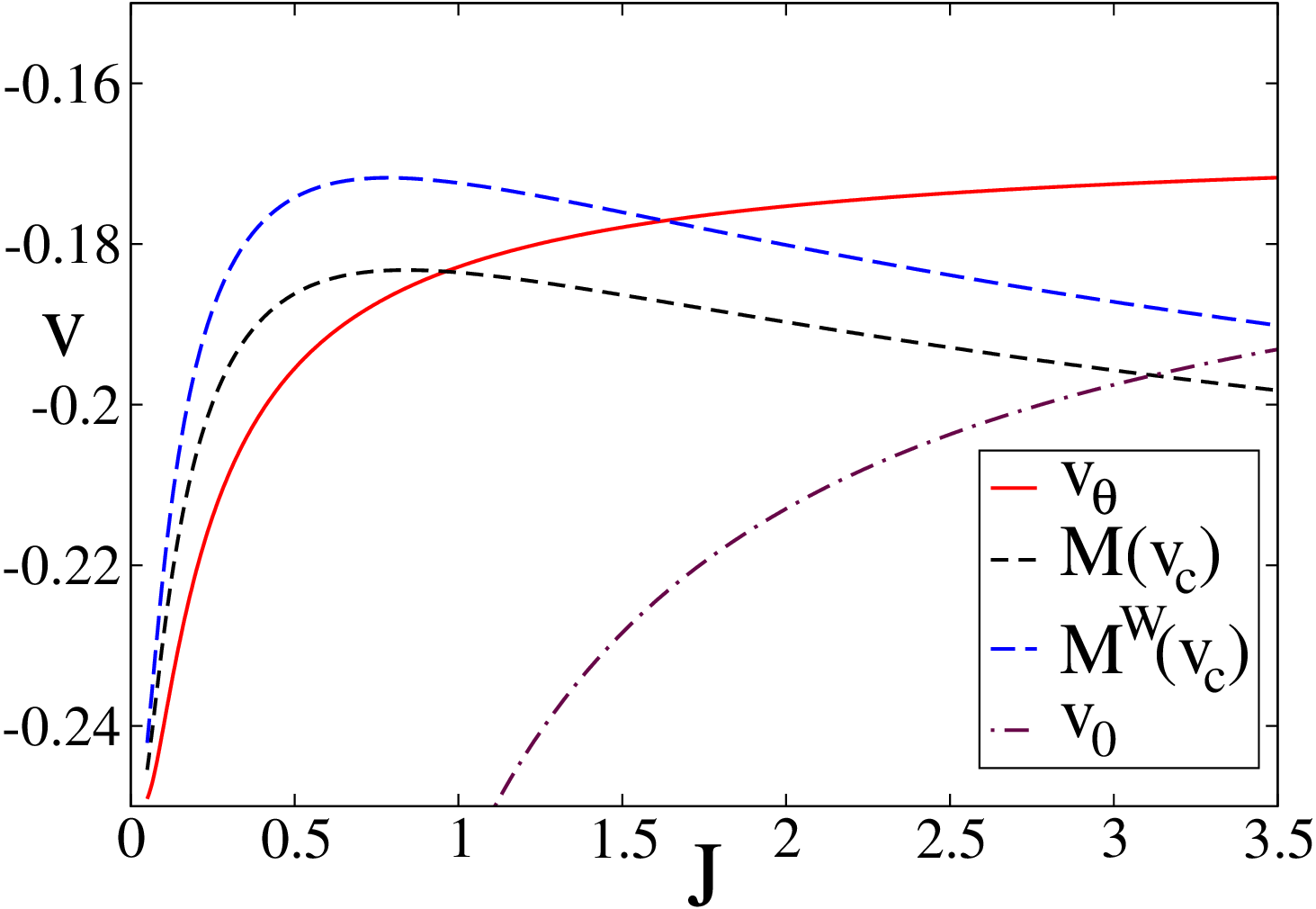}
\caption{The energy $v_\theta$ (full line), the energy ${\cal M}(v_c)$
of the closest saddles at temperature
$T_c$ (lower dashed line) and the energy ${\cal M}^W(v_c)$
of the saddles obtained by the minimization of $W$ starting from an
initial equilibrium configuration at temperature $T_c$ (upper dashed line) as functions of $J$.
The singularity $v_0$ of $\sigma(v)$ is also reported (dot-dashed line).
}
\label{vcj2}
\end{figure}
%%%%%%%%%%%%%%%%%%%%%%%%%%%%%%%%%%%%%%%%%%%%%%

As discussed in the introduction of this paper and in
Ref.~\cite{noiptrig}, the notion of ``inherent saddles'' can {\it
a priori} depend on the way one defines the relation between an
equilibrium configuration and the corresponding stationary point.
We can try to examine what happens if we consider the definition
{(\it 1)} for inherent saddles, defining a new map $v_s={\cal
M}^W(v)$, at least in the paramagnetic phase where the calculation
is straightforward. In this phase $m=0$ and ${\cal
H}(\phi)=h(\phi)$. Thus the spins behave as if they were non
interacting ($J=0$). Thus, for $T>T_c$,

\bea
&W(\phi) = |\nabla H|^2 = \sum_i \left| \frac{\partial h}{\partial \phi} \right|^2  \\
&= \sum_i | \phi_i^3 - \phi_i |^2 =  \sum_i \phi_i^2 ( \phi_i^2 -
1 )^2 \nonumber \eea

The minimization of $W$ can be performed independently for each
degree of freedom; the initial configuration $\phi$ is mapped in a
configuration $\phi^s$ such that
\begin{equation}
\phi^s_i = \begin{cases} &1 \hspace{5pt} \text{  if  }\hspace{5pt} \phi_i \geq \frac{1}{\sqrt{3}} \\
&0\hspace{5pt} \text{  if } \hspace{5pt} \phi_i \in \left[-\frac{1}{\sqrt{3}},\frac{1}{\sqrt{3}}\right] \\
&-1 \hspace{5pt}\text{  if  }\hspace{5pt} \phi_i \leq -\frac{1}{\sqrt{3}}
\end{cases}
\end{equation}
Recalling that $h(0)=0$ and $h(\pm1)=-1/4$, we get (for $T>T_c$):
\bea
v_s^W(T) =
 -\frac{1}{4} \left(1 - \frac{\int_{-1/\sqrt{3}}^{1/\sqrt{3}} d\phi \ e^{-\beta h(\phi)}}
{\int_{-\infty}^{\infty} d\phi \ e^{-\beta h(\phi)}} \right)
\eea

that differs slightly from the expression obtained in the previous
section [Eq.~(\ref{vsTcdist})] where the definition {(\it 2)} of
inherent saddles were used, as the interval of integration is
different. Thus, the energy of the saddles sampled at temperature
$T$ depends slightly, in this model, on the definition of closest
saddles to an equilibrium configurations, i.e. in the way one
defines the basins of attraction of the saddles. In
Fig.~\ref{vcj2} we report, in an expanded scale with respect to
Fig.~7, ${\cal M}^W(v_c)$ together with ${\cal M}(T_c)$ and with
the singularities of the configurational entropy, $v_\theta$ and
$v_0$. We see that the difference between ${\cal M}(v_c)$ and
${\cal M}^W(v_c)$ is of the same order of the difference between
${\cal M}(v_c)$ and $v_\theta$. We conclude therefore that the
relation ${\cal M}(v_c) \sim v_\theta$ holds within the
approximations involved in the calculation of ${\cal M}(v_c)$.

\section{Conclusions}

We characterized the properties of the stationary points of the potential energy surface of the $\phi^4$
model and we compared them with the thermodynamical properties.
We found that the singularity that is observed in the configurational entropy --not in the Euler
characteristic-- is located at an energy $v_\theta$ that is very close
to the energy of the stationary points sampled by the system around the phase transition,
${\cal M}(v_c)$; we got then the relation ${\cal M}(v_c) \sim v_\theta$.
In the previously investigated mean field models \cite{PettiniXY,noieulero,noiptrig}
it was found that ${\cal M}(v_c)=v_c$ and that $v_\theta=v_c$;
our result can be thought as a generalization of the latter relation to the cases where the map
${\cal M}$ is not equal to the identity at $T_c$.
However some uncertainities in the determination of both $v_\theta$ and ${\cal M}(v_c)$ are present.
Indeed, {\it i)} $v_\theta$ is not a true topological singularity as it comes, in our analysis,
from the configurational entropy that is not a topological invariant property of the energy surface;
one should look at the Euler characteristic \cite{pettinisumma} that is however very difficult to determine
in the present model
due to strong cancellations between different saddle orders; and {\it ii)} the exact value of ${\cal M}(v_c)$
has been shown to be slightly dependent on the way one associates to each configuration
the corresponding inherent saddle; in particular, we showed that two different definitions of the inherent
saddle give slightly
different results, and that the difference is of the order of the difference between
${\cal M}(v_c)$ and $v_\theta$.

The possible existence of a singularity in $\chi(v)$ at the critical
energy $v_c$ in the $\phi^4$ mean field model still remains an open problem that needs further investigation.
Moreover, it seems that both the operative definitions of inherent saddle that have been used in the
literature are unable to produce the expected relation ${\cal M}(v_c)=v_\theta$ exactly,
even in such a simple model.
Thus, from the present example one should conclude that the analysis of the thermodynamics
in terms of the stationary points of the potential energy
must be considered an useful but {\it approximate} tool, that has to
be carefully used, evaluating case-by-case the domain of applicability of the method and the approximations
that are necessarily involved.

\acknowledgements

During the final stage of this work, we have been aware of a work on the same subject by
D.~A.~Garanin, R.~Schilling and A.~Scala~\cite{GSS}. We thank them for sending us the preprint before
its publication and for comments on our paper.
We thank F.Sciortino and R.Franzosi for many useful comments and suggestions, and
L.Casetti and M.Pettini for pointing out Ref.~\cite{tesiBaroni}.

%%%%%%%%%%%%%%%%%%%%%%%%%%%%%%%%%%%%%%%%%%%%%%%%%%%%%%%%%%%%%%%%%%%%%%%%%%%%%%%%%%%%%%%%%%%%%%%%%%%%%%%%%%%%%%%%%%%%%%%%%%%%%%%%%%%%%%%%%%%%%%%%%%%%%%%%%%%%%

\appendix

\section{Closest saddles to equilibrium configurations}
\label{app_distance}

\noindent
In this section we will derive the result presented in section \ref{sellevicine}. We have to compute the quantity
\begin{eqnarray}
\nonumber
&\sigma(T;v_s,d)=\frac{1}{N} \int d\phi_i \frac{e^{-\beta H(\phi)}}{Z(T)} \log \int d \psi_i \ \delta(H(\psi)-Nv_s)
\\ &\delta(\partial_i H(\psi)) \ |\det {\rm H}(\psi)| \ \delta \left( d^2 - d^2(\phi,\psi) \right)
\end{eqnarray}
where $d^2(\phi,\psi)=N^{-1} \sum_i ( \phi_i - \psi_i)^2$.
To do that, we need to prove a general relation. Suppose we want to calculate at the saddle point a quantity $Q$ of the form
\begin{eqnarray}
&Q=\frac{1}{N} \int d\phi_i \frac{e^{-\beta H(\phi)}}{Z(T)} \log A(\phi) \nonumber\\
& = \lim_{n \rightarrow 0} \frac{1}{Nn}\left[ \int d\phi_i \frac{e^{-\beta H(\phi)}}{Z(T)} A^n(\phi) - 1 \right] \nonumber \\
&= \lim_{n \rightarrow 0} \frac{1}{Nn} \log \int d\phi_i \frac{e^{-\beta H(\phi)}}{Z(T)} A^n(\phi)
\end{eqnarray}
where we used the relations $\log x = \lim_{n \rightarrow 0} \frac{x^n -1}{n}$ and $\lim_{n \rightarrow 0} (f(n)-1) = \lim_{n \rightarrow 0} \log f(n)$ if $f(n) \rightarrow_{n \rightarrow 0} 1$.
In mean field models, the energy is of the form $H(\phi)=\sum_i h(\phi_i) + N e(m(\phi))$,
where $Nm(\phi)=\sum_i m(\phi_i)$ (in our model, $m(\phi_i)=\phi_i$). Then we have
\begin{eqnarray}
&Q= \lim_{n \rightarrow 0} \frac{1}{Nn} \log \int dm \frac{e^{-\beta N e(m)}}{Z(T)} \nonumber \\
& \times \int D\phi_i \ \delta(m-m(\phi)) \ A^n(\phi) \nonumber \\
&=\lim_{n \rightarrow 0} \frac{1}{Nn} \log \int dm d\hat{m} \frac{e^{-\beta N e(m)}}{Z(T)} \nonumber \\
& \times \int D\phi_i \ e^{ i\hat{m} (N m-\sum_i m(\phi_i))} \ A^n(\phi) \nonumber \\
&=\lim_{n \rightarrow 0} \frac{1}{Nn} \log \frac{1}{Z(T)} \int dm d\hat{m} \ e^{-\beta N (e(m) -T s(n;m,i\hat{m}))}
 \nonumber
\end{eqnarray}
where we defined $D\phi_i = d\phi_i \exp(-\beta h(\phi_i))$ and
\begin{equation} \nonumber
s(n;m,i\hat{m}) = m \ i\hat{m} + \frac{1}{N} \log  \int D\phi_i \ e^{- i\hat{m}  \sum_i m(\phi_i)} \ A^n(\phi)
\end{equation}
Clearly $s(0;m,i\hat{m})$ is the entropic contribution to the free energy as a function of $m$, $\hat{m}$ that we obtain in the calculation of the partition function $Z(T)$, so that
\bea
&  \nonumber f(T)=-\frac{1}{\beta N} \log Z(T) = \min_{m,\hat{m}} [ e(m) - T s(0;m,i\hat{m}) ] \\
&  = e(\mu) - T s(0;\mu,\hat{\mu})=f(0;\mu,\hat{\mu})
\eea
where $(\mu(T),\hat{\mu}(T))$ is the (T-dependent) thermodynamic minimum of the free energy (note that at the saddle point $i\hat{m}=\hat{\mu}$).
Then we have
\begin{equation}
Q= \lim_{n \rightarrow 0} \frac{1}{Nn} \log \int dm \ e^{-\beta N [ f(n;m,i\hat{m}) - f(0;\mu,\hat{\mu}) ]}
\end{equation}
We can now expand $m = \mu + n \mu^{(1)} + o(n^2)$, $i\hat{m} = \hat{\mu} + n \hat{\mu}^{(1)} + o(n^2)$ and
\begin{eqnarray}
& f(n;m,i\hat{m}) - f(0;\mu,\hat{\mu}) =
\frac{\partial f}{\partial m}(0;\mu,\hat{\mu}) \ n \mu^{(1)} \nonumber\\
& + \frac{\partial f}{\partial i\hat{m}}(0;\mu,\hat{\mu}) \ n \hat{\mu}^{(1)} +
\frac{\partial f}{\partial n}(0;\mu,\hat{\mu}) \ n + o(n^2) \nonumber \\
&= \frac{\partial f}{\partial n}(0;\mu,\hat{\mu}) \ n + o(n^2)
\end{eqnarray}
because by definition of ($\mu$,$\hat{\mu}$) we have $\frac{\partial f}{\partial m}(0;\mu,\hat{\mu})=0$, $\frac{\partial f}{\partial i\hat{m}}(0;\mu,\hat{\mu})=0$. We get then the final result:
\begin{equation}
Q= -\beta \frac{\partial f}{\partial n}(0;\mu,\hat{\mu}) =  \frac{\partial s}{\partial n}(0;\mu,\hat{\mu}) \ .
\end{equation}
We have then to calculate (neglecting the term $\mu \hat{\mu}$ that vanish on taking the derivative with respect to $n$):
\begin{eqnarray}
& s(n;\hat{\mu},v_s,d) = \frac{1}{N} \log  \int D\phi_i \ e^{-\sum_i \hat{\mu} \phi_i}
 \prod_{a=1}^n \int d \psi^a_i \times \nonumber \\
& \times \ \delta(H(\psi^a)-Nv_s) \ \delta(\partial_i H(\psi^a)) \times \\
& \times |\det {\rm H}(\psi^a)| \ \delta \left( d^2 - d^2(\phi,\psi^a) \right) \nonumber
\end{eqnarray}
where from the thermodynamic calculation (see section~\ref{sec:thermo}) $\hat{\mu}(T)=-\beta J \mu(T)$ and $\mu(T)$ is given by given by Eq.~\ref{magnetiz}.
We will now neglect the modulus of the determinant of the Hessian matrix replacing in the latter expression
$|\det {\rm H}(\psi^a)|$ with $\det {\rm H}(\psi^a)$, in order to represent the determinant as an integral over
fermionic variables. We will see later how to restore the correct sign in this term.
Using a superfield representation \cite{noiptrig} we get
\begin{eqnarray}
& s(n;\mu,v_s,d) = \frac{1}{N} \log  \int D\phi_i \ e^{\beta J \mu \sum_i \phi_i}
\prod_{a=1}^n \int \frac{d \gamma_a}{2\pi} e^{N \gamma_a v_s} \nonumber \\
& \times \int {\cal D}\Psi^a_i \  \exp \left[ \int d\bar{\theta} d\theta (1-\gamma_a \theta \bar{\theta}) H(\Psi^a) \right] \times \\ & \times \delta \left( N d^2 - \sum_i (\phi_i - \psi_i^a)^2 \right) \nonumber
\end{eqnarray}
We will now: {\it i)} substitute the expression $H(\Psi^a)=\sum_i h(\Psi^a_i) - J N m(\Psi^a)^2/2$; {\it ii)} insert some $\delta$-functions for $m_a=m(\Psi^a)$ and the corresponding integral representation with a multiplier $\hat{m}_a$; {\it iii)} neglect all the product and sum signs related to the index $a$; {\it iv)} use the integral representation for the $\delta$-function of $d^2$ with a multiplier $\lambda_a$. Then we get an expression that has to be maximized with respect to all the parameters to get the saddle point value of $s(n;\mu,v_s,d)$:
\begin{eqnarray}
& s(n;\mu,v_s,d)=\max_{\rm all par} \Big[ \sum_a \gamma_a v_s - \nonumber \\
& \sum_a \int d\bar{\theta} d \theta [ (1- \gamma_a \theta \bar{\theta}) \frac{J m^2_a}{2} - m_a \hat{m}_a ] \nonumber \\
& + \sum_a \lambda_a d^2 + \log {\cal S}(\mu,\hat{m}_a,\gamma_a,\lambda_a) \Big] \\
& {\cal S}(\mu,\hat{m}_a,\gamma_a,\lambda_a) = \int d\phi \ {\cal D}\Psi^a \exp \Big[ -\beta {\cal H}(\phi) \nonumber \\
&+ \sum_a \int d\bar{\theta} d\theta (1-\gamma_a \theta \bar{\theta}) h(\Psi^a)  \nonumber \\
& - \sum_a \int d\bar{\theta} d \theta \ \hat{m}_a \Psi^a - \sum_a \lambda_a (\phi-\psi^a)^2 \Big] \nonumber
\end{eqnarray}
where ${\cal H}(\phi) = h(\phi) - J \mu \phi$ as in Eq.~\ref{singleparticle}.
As usual, we will assume that: {\it i)} there is symmetry between the replicas ($m_a=m$, etc.); {\it ii)} all the fermionic components vanish at the saddle point. Then we get
\begin{eqnarray}
& s(n;\mu,v_s,d)=\max_{\rm all par} \Big[ n \Big( \gamma (v_s + \frac{J m_0^2}{2}) - J m_0 m_3 \nonumber \\
& + \hat{m}_0 m_3 + \hat{m}_3 m_0 + \lambda d^2 \Big) + \log {\cal S}(\mu,\hat{m},\gamma,\lambda) \Big]
 \nonumber \\
& {\cal S}(\mu,\hat{m},\gamma,\lambda) = \int d\phi \ e^{ -\beta {\cal H}(\phi) }
 \Big[ \int {\cal D}\Psi \exp \Big( \int d\bar{\theta} d \theta \nonumber \\
& [ (1 - \gamma \theta \bar{\theta}) h(\Psi) - \hat{m} \Psi]  -  \lambda (\phi-\psi)^2 \Big) \Big]^n \nonumber
\end{eqnarray}
Now we have to take the derivative of $s$ with respect to $n$ at $n=0$. By direct computation
\begin{eqnarray}
&\sigma(\mu;v_s,d)=\max_{\rm all par} \frac{\partial s}{\partial n}(0;\mu,v_s,d) = \nonumber \\
&\max_{\rm all par} \Big[ \Big(  \gamma (v_s + \frac{J m_0}{2})
 - J m_0 m_3  + \\
& \hat{m}_0 m_3 + \hat{m}_3 m_0 + \lambda d^2 \Big)
 + \int d \phi \frac{e^{-\beta {\cal H}(\phi)}}{{\cal Z}(T)}
\log  \int {\cal D}\Psi \nonumber \times \\
& \times \exp \Big( \int d\bar{\theta} d \theta \ [ (1 - \gamma \theta \bar{\theta}) h(\Psi) - \hat{m} \Psi]
 -  \lambda (\phi-\psi)^2 \Big) \nonumber
\end{eqnarray}
Some of the parameters can be easily eliminated computing the derivatives of $\sigma$; defining $\alpha=Jm_0$
one is left with the following expression:
\bea
 \label{sigmaselleAPP}
&\sigma(\mu;v_s,d)=\max_{\alpha,\gamma,\lambda} \Big[ \gamma \Big( v_s - \frac{\alpha^2}{2 J} \Big) \\
& + \lambda d^2
+ \int d\phi \ {\cal P}(\phi) \ \log \sum_\xi e^{-\gamma {\cal H}(\phi_\xi(\alpha)) -
\lambda (\phi - \phi_\xi(\alpha))^2} \Big]
\nonumber
\eea
where ${\cal P}(\phi)$ is defined in Eq.~\ref{singleparticle}, $\xi=(-,0,+)$ and the $\phi_\xi(\alpha)$
are defined in Eq.~\ref{solu3grado}.
Note that in the latter expression the term $\xi = 0$ in the logarithm should have a minus sign:
this is a consequence of the absence of the modulus of the determinant of the Hessian matrix that we neglected
above. Taking the modulus into account corresponds to neglecting the minus sign of the term $\xi=0$.
Performing the derivatives with respect to $\alpha$,$\gamma$ and $\lambda$ one obtains
the following equations:
\bea
\nonumber
&\alpha = J \int d\phi \ {\cal P}(\phi) \sum_\xi {\cal P}_\xi(\phi,\gamma,\lambda) \ \phi_\xi(\alpha) \\
\label{APP1}
&d^2 =  \int d\phi \ {\cal P}(\phi) \sum_\xi {\cal P}_\xi(\phi,\gamma,\lambda) \ (\phi - \phi_\xi(\alpha))^2 \\
&v_s = \frac{\alpha^2}{2J} +
\int d\phi \ {\cal P}(\phi) \sum_\xi {\cal P}_\xi(\phi,\gamma,\lambda) \ {\cal H}(\phi_\xi(\alpha))
\nonumber
\eea
where
\begin{equation}
{\cal P}_\xi(\phi,\gamma,\lambda) =
\frac{e^{-\gamma {\cal H}(\phi_\xi(\alpha)) - \lambda (\phi -\phi_\xi(\alpha))^2}}
{\sum_\xi e^{-\gamma {\cal H}(\phi_\xi(\alpha)) - \lambda (\phi -\phi_\xi(\alpha))^2}}
\end{equation}
We want now to minimize $d^2$ with the condition $\sigma = 0$. It is easy to show from Eq.~(\ref{APP1}) that
$\frac{\partial d^2}{\partial \lambda} \leq 0$. Then we expect that the minimum distance is obtained in the
$\lambda \rightarrow \infty$ limit (see Ref.~\cite{noiptrig} for a detailed discussion of this point).
It is easy to see that
\begin{equation}
\lim_{\lambda \rightarrow \infty} {\cal P}_\xi(\phi,\gamma,\lambda) = \chi^\alpha_\xi(\phi)
\end{equation}
where the function $\chi^\alpha_\xi(\phi)$ is equal to 1 if $\phi_\xi(\alpha)$ is the closest to $\phi$
and 0 otherwise.
Thus, if we define $\tilde\phi(\phi,\alpha)$ as
\begin{equation}
\tilde\phi(\phi,\alpha)=\sum_\xi \chi_\xi(\phi) \phi_\xi(\alpha)
\end{equation}
(i.e. $\tilde\phi$ is the closest $\phi_\xi$ to $\phi$),
in the $\lambda \rightarrow \infty$ Eq.s~(\ref{APP1}) become
\bea
\nonumber
&\alpha = J \int d\phi \ {\cal P}(\phi) \ \tilde\phi(\phi,\alpha) \\
\label{APP2}
&d^2 =  \int d\phi \ {\cal P}(\phi) \ (\phi - \tilde\phi(\phi,\alpha))^2 \\
&v_s = \frac{\alpha^2}{2J} +
\int d\phi \ {\cal P}(\phi) \ {\cal H}(\tilde\phi(\phi,\alpha))
\nonumber
\eea
The first of the latter equation has to be interpreted as a self-consistency equation that gives the value
of the magnetization of the closest saddles to the equilibrium configurations, $\alpha_s(T)$. The second
and third equations give the average distance $d^2(T)$ and the average potential energy $v_s(T)$.
Finally, observing that
\bea
&\lim_{\lambda \rightarrow \infty} \log \sum_\xi e^{-\gamma {\cal H}(\phi_\xi(\alpha)) -
\lambda (\phi - \phi_\xi(\alpha))^2} =\\
&= -\gamma {\cal H}(\tilde\phi(\phi,\alpha)) -
\lambda (\phi - \tilde\phi(\phi,\alpha))^2
\eea
substituting the latter expression in Eq.~(\ref{sigmaselleAPP}) and using Eq.s~\ref{APP2} one obtains
\begin{equation}
\lim_{\lambda \rightarrow \infty} \sigma = 0
\end{equation}
consistently with our initial assumption.

%%%%%%%%%%%%%%%%%%%%%%%%%%%%%%%%%%%%%%%%%%%%%%%%%%%%%%%%%%%%%%%%%%%%%%%%%%%
%                             REFERENCES
%%%%%%%%%%%%%%%%%%%%%%%%%%%%%%%%%%%%%%%%%%%%%%%%%%%%%%%%%%%%%%%%%%%%%%%%%%%

%%%%%%%%%%%%%%%%%%%%%%%%%%%%%%%%%%%%%%%%%%%%%%%%%%%%%%%%%%%%%%%%%%%%%%%%%%%%%%%%%%%%%%%%%%%%%%
%%%%%%%%%%%%%%%%%%%%%%%%%% FIGURES %%%%%%%%%%%%%%%%%%%%%%%%%%%%%%%%%%%%%%%%%%%%%%%%%%%%%%%%%%
%%%%%%%%%%%%%%%%%%%%%%%%%%%%%%%%%%%%%%%%%%%%%%%%%%%%%%%%%%%%%%%%%%%%%%%%%%%%%%%%%%%%%%%%%%

\end{document}